\documentclass[twocolumn,showpacs,aps]{revtex4}
\usepackage{bm}
\usepackage{graphicx}
\usepackage{bm}

\begin{document}

\title{Electron pairing: from metastable electron pair to bipolaron}

\author{G.-Q. Hai$^{1}$\footnote{Corresponding e-mail: hai@ifsc.usp.br}, L. C{\^a}ndido$^2$,
B. G. A. Brito$^{3}$, and F. M. Peeters$^{4}$}

\affiliation{$^{1}$Instituto de F\'{\i}sica de S\~{a}o Carlos, Universidade de S\~{a}o Paulo, 13560-970,
S\~{a}o Carlos, SP, Brazil}
\affiliation{$^{2}$Instituto de F\'isica, Universidade Federal de Goi\'as, 74001-970, Goi\^ania, GO, Brazil}
\affiliation{$^{3}$Departamento de F\'isica, Instituto de Ci\^encias Exatas e Naturais e Educa\c{c}\~ao (ICENE),
Universidade Federal do Tri\^angulo Mineiro - UFTM, 38064-200, Uberaba, MG, Brazil}
\affiliation{$^{4}$Departement Fysica, Universiteit Antwerpen, Groenenborgerlaan 171, 2020 Antwerpen, Belgium}

\begin{abstract}
Starting from the shell structure in atoms and the significant correlation within electron pairs, we
distinguish the exchange-correlation effects between two electrons of opposite spins occupying
the same orbital from the average correlation among many electrons in a crystal.
In the periodic potential of the crystal with lattice constant larger than the effective Bohr radius
of the valence electrons, these correlated electron pairs can form a metastable energy band above the
corresponding single-electron band separated by an energy gap.
In order to determine if these metastable electron pairs can be stabilized, we calculate the many-electron
exchange-correlation renormalization and the polaron correction to the two-band system with single
electrons  and electron pairs. We find that the electron-phonon interaction is essential to counterbalance
the Coulomb repulsion and to stabilize the electron pairs. The interplay of the electron-electron
and electron-phonon interactions, manifested in the exchange-correlation energies, polaron effects, and screening,
is responsible for the formation of electron pairs (bipolarons) that are located on the Fermi surface
of the single-electron band.
\end{abstract}

\maketitle

\section{Introduction}

Since the discovery of high $T_c$ superconductivity by Bednorz and M{\"u}ller\cite{BM} in 1986,
great progresses have been made in the experimental and theoretical investigation of unconventional
superconductivity.
However, the mechanism of electron pairing in unconventional superconductors remains one
of the most challenging and unresolved problems in condensed matter physics.\cite{Norman,BB17,JZ16}
Vast experimental evidences have shown that electron pairing and unconventional superconductivity
occur in many different materials, such as cuprates,\cite{BM,Norman,BB17,JZ16} iron-based
superconductors,\cite{hosono,sprau17,kasahara,gerber17} and carbon-based
superconductors,\cite{kubozono,heguri} etc. Although there are many different theories for
unconventional superconductivity, almost all theories follow
the basic idea of the BCS theory\cite{bcs}. They presume that there is some
effective attraction between electrons leading to Cooper pairing which spontaneously condense into
a collective non-Fermi liquid state.

We would like to mention some very recent experimental results related to the electron-pairing
mechanism in unconventional superconductors. Bo$\check{\rm z}$ovi$\acute{\rm c}$ et al. reported
very impressive and accurate results on superconductivity in high-$T_c$ cuprates.\cite{bozovic16}
They synthesized atomically perfect thin films and multilayers of cuprates La$_{2−x}$Sr$_x$CuO$_4$ (LSCO)
and measured the absolute value of the magnetic penetration depth and the phase stiffness with high
accuracy in thousands of samples. The large statistics revealed clear trends in the intrinsic properties
of the cuprate superconductors. They found that the obtained results disagree with the BCS theory
in any variant, i.e. clean or dirty, including the Migdal-Eliashberg theory. Rather, the experimental data
indicated small (local) and very light electron pairs with mass on the order of an electron mass.
These pairs are preformed well above $T_c$ and at $T_c$ undergo Bose-Einstein
condensation.\cite{bozovic16,bozovic17} Investigations performed by Zhong et al.\cite{zhong} and by
Ren et al.\cite{ren16} challenged the $d$-wave pairing mechanism in cuprates.
With scanning tunneling spectroscopy, they revealed anisotropic and nodeless superconducting gaps
in the cuprate superconductors Bi$_{2}$Sr$_{2}$CaCu$_{2}$O$_{8+\delta}$ (Bi2212) and YBa$_2$Cu$_3$O$_{7-x}$ (YBCO).
In their paper, Zhong et al.\cite{zhong} affirmed that ``this is contradictory to the nodal $d$-wave pairing
scenario that is often thought to be the most important result in the 30-year study of the HTS
mechanism of cuprates".

Important progresses in recent investigations on the electron pairing mechanism in iron-based
superconductors indicate small and preformed Cooper pairs.\cite{sprau17,kasahara,gerber17}
For instance, using Bogoliubov quasiparticle interference imaging, Sprau et al.\cite{sprau17}
found that the superconducting energy gap in FeSe is extremely anisotropic and nodeless. Their
investigation discovered the existence of orbital-selective Cooper pairing in FeSe.
Gerber et al.\cite{gerber17} combined two time-domain experiments
into a “coherent lock-in” measurement in the terahertz regime and was able to quantify the electron-phonon
coupling strength in FeSe. Their study revealed a strong enhancement of the electron-phonon coupling
strength in FeSe owing to electron correlations and highlighted the importance of the
cooperative interplay between electron-electron and electron-phonon interactions

In this paper, we present a theory for electron pairing in crystals where we consider the electron-electron
correlation, the periodic potential of the crystal lattice, and the electron-phonon interaction.
Our theory is different from all previous ones. We obtain preformed small electron pairs
in the periodic potential of the crystal. This is essentially an orbital dependent electron-pairing theory
in which the exchange-correlation between two electrons occupying the same orbital is decisive for
pair formation. However such pairs are metastable unless the electron-phonon interaction is included.
The calculations show that the electron-phonon coupling and the polaron effects are responsible
for the stabilization of the electron pairs.

Our basic idea for the electron pairing process is that the electron-electron correlation
is orbital dependent. Within the jellium model for the electron gas in solids,\cite{Mahan}
the atomic nuclei that form the periodic lattice are smeared out into a uniform positive charge distribution.
Each electron is totally delocalized. Therefore, many electrons ``see" each other with their fluctuation
potential at the same time and thus correlate all at once, giving rise to collective screening and
oscillation effects. However, in atoms and molecules, significant correlations occur within electron
pairs.\cite{OS,slupski,hattig} Strong exchange-correlation interaction between two electrons in the same
orbital manifests in the shell structure of atoms and also in the covalent and ionic bonding in molecules.
Our starting point in this study is to distinguish the exchange-correlation effects between two electrons
of opposite spins occupying the same atomic orbital
from the average correlation among many electrons in a crystal. This may happen in a crystal
but the electrons have to ``feel" the nuclei potential well. This leads to a preliminary condition
that the effective Bohr radius of the valence electrons in the crystal has to be comparable
or smaller than the lattice constant. For instance, for a cuprate crystal with effective electron mass
$m^* \simeq 5m_0$ and static dielectric constant $\epsilon_0 \simeq 30$, the effective Bohr radius
a$_B \simeq 3.2$ \AA\ is smaller than the lattice constant of about 3.8 \AA.
Because we want to show that the electron-pair correlation in atoms can manifest themselves in
electron transport in crystals, our calculations have to start first with the formation
of energy bands.\cite{Ashcroft}

In order to find out the electron-pair states in the crystal, we will first establish a simple
crystal model to discuss the physical process. We consider a ``hydrogen solid" model with single-electron
state of H atom and electron-pair state of the H$^-$ ion. We will show that, besides the energy bands from
the single-electron energy levels of individual atoms in the crystal, there can exist
a metastable electron-pair energy band from the correlated electron pairs of the H$^-$ state
for the lattice constant $\lambda$ being larger than the effective Bohr radius a$_B$.\cite{hai2014}
The electron pairs are metastable because the Coulomb repulsion is strong overwhelming
the exchange-correlation. In order to stabilize them we have to include the electron-phonon
interaction to counterbalance the Coulomb repulsion. Therefore, the electron-phonon interaction
is necessitated in a natural way in the electron pairing process.

This paper is organized as follows. In Sec. II we calculate and discuss the metastable electron-pair
energy band in two-dimensional square-lattice crystals. In Sec. III we present the many-particle
Hamiltonian consisting of electrons in both the single-electron and electron-pair bands being coupled to
the LO-phonons. In Sec. IV and Sec. V, we calculate the many-particle
exchange-correlation (xc) energies due to electron-electron interactions and polaron energies
due to electron-phonon interactions. Then, we show in Sec. VI the conditions under which the metastable
electron pairs can be stabilized in the ground state including many-body effects.
Finally, we summarize our work in Sec. VII.

In the calculations, we will use effective Bohr radius ${\rm a}_B = \epsilon_0 \hbar^2 /m^* e^2 $
and effective Rydberg ${\rm R}_y=\hbar^2/2m^* {\rm a}^2_{B}$ as the units for length and energy,
respectively.

\section{Metastable electron pairs in crystal}

Electronic band structure is of fundamental importance for our understanding of many physical properties
of solids. Within the independent electron approximation, the electron states are of Bloch form in the
periodic potential of the crystal lattice. The effects of electron-electron interactions are accounted
for by an effective potential which repeats this periodicity.\cite{Ashcroft} In this section, we will
show that, besides the energy bands from the single-electron states there can exist a metastable
electron-pair energy band depending upon the crystal structure and potential. This metastable electron-pair
band originates from two correlated electrons of opposite spins occupying the same atomic orbital.

\subsection{Two-electron atoms}

Our study will start with the simplest electron-pair system, i.e., two-electron atoms. These helium-like
atoms with two electrons of opposite spins occupying the same orbital, e.g. helium atom He and negatively
charged hydrogen anion H$^-$ have played an important role in the development of theoretical physics
in the last century.\cite{Tanner} It is a challenge to determine accurately the correlation energy,
even in simple systems such as He atom and H$^-$ ion.\cite{Tanner,Rau,HH,Loos} Hylleraas' result for
the ground-state energy of He atom obtained in 1929 was -5.80648 R$_y$\cite{Hylleraas1929}.
After generations of calculation, very accurate (non-relativistic) ground-state
energies\cite{Kleindienst,Baker,Thakkar,Frolov,Nakashima,Turbiner} of two electron atoms have been
obtained: -5.807448754068$\cdots$ R$_y$ for He and -1.055502033088$\cdots $R$_y$ for H$^-$.
Recently, using high-precision variational calculations Estienne et al.\cite{Estienne} determined the critical
nuclear charge $Z_c$=0.911 028 224 077 255 73(4) which is the minimum charge required to bind
two electrons in a helium-like atom.

On the other hand, the famous experiment on two-electron atoms by Madden and Codling\cite{Madden}
revealed that the simple model based on independent particle picture is inappropriate to characterize
a series of doubly excited states because of strong electron-electron correlation.\cite{Tanner,Madden}
In comparison with the single-electron states of the H atom, the H$^-$ ion is a closed-shell system
with two strongly correlated electrons. Such an electron-pair state is different in its nature from
the single-electron states because of the strong correlation. It should be recognized
as a new strongly correlated electronic state.

Negative hydrogen ion H$^-$ in two-dimensional (2D) system has also been investigated
in the last decades mostly because of the discovery of its counterpart D$^-$ center in
2D semiconductor quantum wells\cite{Huant}. The D$^-$ center is a negatively charged shallow donor
impurity center in semiconductors, such as a negatively charged Si impurity in a GaAs quantum well.
It is an H$^-$-like state in solid-state environment but with very different energy and length scales
(e.g., in GaAs, the effective Rydberg R$_y$=5.9 meV and effective Bohr radius a$_B$=98 \AA).
Therefore, the D$^-$ center in semiconductors is considered as an ideal ``laboratory" to
study the H$^-$ properties, for instance, in high magnetic fields.\cite{Shi}
Earlier variational calculation by Phelps and Bajaj found the energy of the H$^-$ in 2D is
-4.48 R$_y$.\cite{Phelps} Further numerical calculations obtained -4.48054 R$_y$
by Ivanov and Schmelcher\cite{Ivanov} and -4.4804798 R$_y$ by Ruan {\it et al.}\cite{Ruan}.

\begin{table}[t!]
\label{tableI}
\caption{The ground-state energies of negative hydrogen ion H$^-$ and hydrogen atom H
in 2D and 3D. E$_b$ is the binding energy of H$^-$. The energy per electron
in H$^-$ state is given by $ \varepsilon_{p} = E_{{\rm H}^-}/2$. Energies are in units of R$_y$.}
\centering
\begin{tabular}{ccccc}
\hline\hline
 &  $E_{{\rm H}^-}$    &   $E_{{\rm H}}$  &  $E_b$ &  $\varepsilon_{p}$ \\
\hline
 3D \hspace{0.5cm} &$-1.055 \;\;\; $ &$-1.0 \;\;\;$ &$ 0.055 \;\; $ &$-0.528 $ \\
 2D \hspace{0.5cm} &$-4.48 \;\;\; $  &$-4.0 \;\;\;$ &$ 0.48  \;\;\;\;$  &$-2.24 $\\
\hline\hline
\end{tabular}
\end{table}

In Table I we compare the ground-state energies of the 2D and 3D H$^-$ states.
The binding energy $E_b$ is defined as the difference between the energies $E_{{\rm H}}$
of the neutral H atom and $E_{{\rm H}^-}$ of the H$^-$ ion. This is the energy required to remove one of
the two electrons from the H$^-$ ion to infinity. It is also called electron
affinity of the hydrogen atom. One sees that the binding energy $E_b$ of the H$^-$ in 2D is almost
9 times larger than that in 3D because electron correlation in 2D is much stronger.
In the last column we also give the energy per electron $\varepsilon_p$ in the H$^-$ state.

\begin{figure}[b!]
            {\includegraphics[scale=0.8]{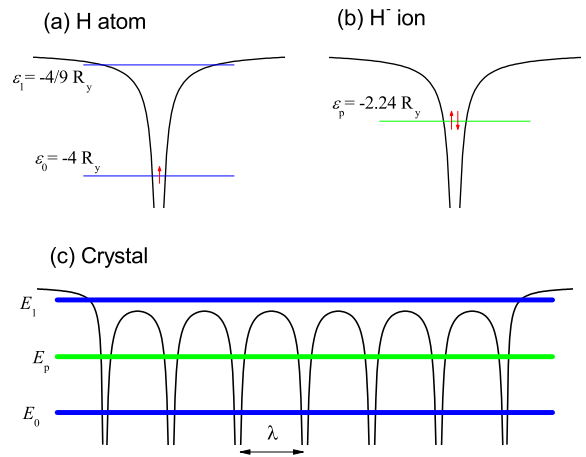}}
       \caption{(a) A 2D H atom, (b) a 2D H$^{-}$ ion, and (c) a ``hydrogen solid".
       The horizontal lines indicate the energy levels and bands.}
       \label{fig1}
\end{figure}

\subsection{``Hydrogen solid" model with both the single-electron and electron-pair states}

In order to explain the so-called Mott insulator and metal-insulator transition\cite{Mott,VD},
Mott considered a hydrogen solid model with the single-electron energy band only, i.e., a
simple cubic lattice crystal of one-electron atoms and made the following discussion\cite{Mott}.
For small values of the lattice constant $\lambda$, there is a half-filled band
in such a crystal and thus it is metallic. If one varies the lattice constant
to large values (but not so large as to prevent tunnelling), the Coulomb interaction $U$
for two electrons occupying the same atomic site overcomes the kinetic energy
(characterized by the band width $W$).
In this case, each electron should be assigned to its parent atom.
The crystal must be nearly the same as
a collection of isolated neutral hydrogen atoms and thus it is an insulator.
This reveals a competition between potential and kinetic effects.
At large $\lambda$ (small $W$) the Coulomb repulsion $U$ dominates, the electrons
are localized, and the system is insulating.\cite{Mott,VD}. The idea of Mott led to
the theoretical model introduced by Hubbard.\cite{Hubbard}
The Hubbard model traces the insulating behavior to strong Coulomb repulsion between electrons
occupying the same orbital. The competition between the kinetic
and Coulomb energies gives rise to strong electron-electron correlations.
The Hubbard model was proposed originally to describe the transition between conducting
and insulating systems.
It has also been widely used to study materials with strongly correlated electrons
and high-temperature superconductivity.\cite{Scalapino}

In this paper we will consider both the single-electron and electron-pair states
in the ``hydrogen solid" model. Our calculation will be performed for 2D systems because
we can obtain more accurate numerical results in 2D. Another reason why 2D systems are more interesting
is that electron-electron correlations are stronger. Many unconventional superconductor materials
are found to be essentially two dimensional.
Fig.~1 shows diagrams representing (a) a 2D H atom and (b) a 2D H$^-$ ion with their respective
energy levels. The nuclear potential $V_a({\bf r}-{\bf R}_m)$ of the atom is represented by the black curves,
where ${\bf R}_m$ is the position of the nucleus. The single-electron levels of a 2D H atom are given by
$\varepsilon_i=- {\rm R}_y /(i+1/2)^{2}$ (for $i$=0,1,2,...).  The energy level of a correlated electron pair
in H$^-$ ion, i.e., the energy per electron in the ground state, is given by
$\varepsilon_p=E_{{\rm H}^-}/2 = -2.24$ R$_y$. We remind that a single H$^-$ ion is stable.

We now consider the following  ``hydrogen solid":
$N$ atoms are arranged into a simple-lattice crystal at positions ${\bf R}_m$ (for $m=0,1,2,\cdots,N-1$)
with lattice constant $\lambda$ of the order of the Bohr radius a$_B$ as indicated in Fig.~1(c).
The crystal potential for an electron at ${\bf r}_j$ is given by
\begin{equation}\label{VCC}
V_c({\bf r}_j) = \sum_{m=0}^{N-1} V_a({\bf r}_j-{\bf R}_m).
\end{equation}
It is known that the single-electron levels $\varepsilon_i$ of individual atoms form energy bands
$E_{i}({\bf k})$ in such a crystal\cite{Ashcroft} as indicated by the horizontal thick-blue lines in Fig.~1(c).
In principle, there is also the possibility that two electrons of opposite spins occupy the same atomic orbital
forming an H$^-$-like state in the crystal, but it becomes unstable due to the presence of the neighbor atoms.
Therefore, the counterpart of the H$^-$ state in a crystal has never been investigated.
In this section we will show that, though such an electron pair
is unstable in a crystal due to the Coulomb repulsion, they may form a metastable energy band
(indicated by the horizontal thick-green line in Fig. 1(c)) depending on the crystal structure and potential.
In such a crystal the lattice constant $\lambda$ should not be so small as to prevent individual atoms
to bind two electrons, but not so large as to prevent co-tunnelling of an electron pair
between the neighbor unit cells.

In the center of mass and relative coordinates ({\bf R},{\bf r}) of the two electrons at
${\bf r}_1$ and ${\bf r}_2$, defined by
\begin{equation}
{\bf R}=\frac{1}{2}({\bf r}_1 + {\bf r}_2) \;\;\; {\rm and}\;\;\;\;
{\bf r}={\bf r}_1 - {\bf r}_2,
\end{equation}
the wavefunction of an individual electron pair with energy 2$\varepsilon_p$ bound to the atom
at ${\bf R}_m$ is given by $\phi({\bf R}-{\bf R}_m,{\bf r})$.
The Schr\"odinger equation for two electrons in the crystal potential given by Eq.~(\ref{VCC})
can be written as,
\begin{eqnarray}\label{SE2EC}
 [&-& {\frac{1}{2}} \nabla^2_{{\bf R}}-2\nabla^2_{{\bf r}} + V_c({\bf r}_1)
+ V_c({\bf r}_2) + \frac{2}{|{\bf r}|} ] \Psi({\bf R},{\bf r}) \nonumber \\
&=& 2 E_{p} \Psi({\bf R},{\bf r}),
\end{eqnarray}
where ${\bf r}_1={\bf R}+\frac{_1}{^2}{\bf r} $ and ${\bf r}_2 ={\bf R}-\frac{_1}{^2}{\bf r}$.
The term $2/|{\bf r}|$ is the Coulomb repulsion potential between the two electrons.
We should bear in mind that, due to electron-electron repulsion, the ground state of this system
can be found for $|{\bf r}_1 - {\bf r}_2|=|{\bf r} | \to \infty $. In other words, the ground state
of this two-electron system corresponds to two non-interacting single electrons separated by an
infinitely long distance. But we are looking for the quantum states of two-correlated electrons
occupying the same orbital in the same unit cell in the crystal with the average separation
$\langle r \rangle$ being less than the lattice constant $\lambda$.  Therefore, the electron-pair
states in the crystal are metastable. The calculations in Sec.~II-C will confirm that such a
metastable state does exist in 2D periodic potentials.

If the electron-pair states of two correlated electrons in the crystal can be approximated by a
{\it linear combination of the electron-pair wavefunctions $\phi({\bf R}-{\bf R}_m,{\bf r} )$
of single atoms}, written as,
\begin{equation}
\Psi({\bf R},{\bf r})= \sum_m c_m \phi({\bf R}-{\bf R}_m,{\bf r} )
\end{equation}
for $\langle r \rangle < \lambda$, we can obtain the following homogeneous linear equations,
\begin{equation}\label{CIRM}
\sum_m \left[ J_p({\bf R}_m-{\bf R}_n)
+ 2 \left(E_p - \varepsilon_{p} \right) \alpha_p({\bf R}_m-{\bf R}_n)\right] c_m =0,
\end{equation}
for $n,m=0,1,2,..., N-1$, where $\alpha_p( {\bf R}_{l} )$ is the overlap integral
\begin{equation}
  \alpha_p({\bf R}_l)=  \int d{\bf R}\int d{\bf r}
 \phi^*({\bf R}-{\bf R}_l,{\bf r}) \phi({\bf R},{\bf r}),
\end{equation}
with $ \alpha_p( {\bf R}_{l}=0 )=1$, and
\begin{equation}
  J_p({\bf R}_l)= - \int d{\bf R} \int d{\bf r} \phi^*({\bf R}-{\bf R}_l,{\bf r})
 \Delta V_l({\bf R}, {\bf r}) \phi({\bf R},{\bf r}),
\end{equation}
with
\begin{equation}
\Delta V_l({\bf R},{\bf r})= \sum_{n\neq l}
\left[ V_a({\bf R}-{\bf R}_n+\frac{_1}{^2}{\bf r}) + V_a({\bf R}-{\bf R}_n-\frac{_1}{^2}{\bf r}) \right] .
\end{equation}

We observe that Eq.~(\ref{CIRM}) for $c_m$ depends only on ${\bf R}_l = {\bf R}_m-{\bf R}_n $.
This is an eigenvalue problem of a block circulant matrix.\cite{DAVIS}
The solution has the following form
\begin{equation}
   c_m = C \cdot e^{i{\bf k} \cdot {\bf R}_m},
\end{equation}
where the vector ${\bf k}$ should be a reduced wavevector in the first Brillouin zone and
$C$ is the normalization constant.
We finally obtain the electron-pair wavefunction in the crystal given by
\begin{equation}
\Psi_{\bf k}({\bf R},{\bf r})
= \frac{\sum_l e^{i{\bf k}\cdot{\bf R}_{l}} \phi({\bf R}-{\bf R}_l, {\bf r})}
{\sqrt{N \left(1+ \sum_{l \neq 0} e^{i{\bf k}\cdot {\bf R}_{l}  } \alpha_p( {\bf R}_{l} )\right)}}.
\end{equation}
The above wavefunction is a Bloch wavefunction in
the center of mass coordinates {\bf R} because it can be written as
\begin{equation}
\Psi_{\bf k}({\bf R},{\bf r}) =  C  e^{i{\bf k}\cdot{\bf R}}
\left[\sum_l e^{-i{\bf k}\cdot({\bf R}-{\bf R}_{l})} \phi({\bf R}-{\bf R}_l, {\bf r})\right],
\end{equation}
where the part in the square brackets is a periodic function
in the coordinates ${\bf R}$ with period of the crystal lattice.
The dispersion relation of the electron-pair band is given by
\begin{equation}
 E_{p}({\bf k})= \varepsilon_{p}- {{\sum_l  J_p({\bf R}_l)  e^{i{\bf k} \cdot {\bf R}_l} }
 \over {2 \sum_l \alpha_p({\bf R}_l) e^{i{\bf k} \cdot {\bf R}_l} }}.
\end{equation}

Because the considered electron-pair wavefunction $\phi({\bf R}-{\bf R}_m, {\bf r})$ of a
two-electron atom has essentially the $s$-symmetry\cite{HH,Phelps}, the dispersion relation
of the electron-pair band
in the 2D square-lattice crystal with lattice constant $\lambda$ can be approximated as
\begin{equation}\label{E2DSQ}
 E_{p}({\bf k})= \varepsilon_{p}- \frac{1}{2} J_p(0)- J_p({\bf R}_1) [ \cos(k_x \lambda) + \cos (k_y \lambda) ],
\end{equation}
where only the nearest-neighbor tunneling term $J_p({\bf R}_1)$ is considered. The value of
$J_p({\bf R}_1)$ determines the bandwidth of the electron-pair band. Because this is a co-tunneling
process of two paired electrons between the neighbor sites and the effective mass of the electron pair
is twice of a single electron, the electron-pair bandwidth
should be much smaller than that of the single-electron band. This dispersion relation will
be confirmed in the next section by making numerical calculations of the metastable electron-pair band
in 2D periodic potential of a square lattice.

\subsection{Metastable electron-pair band in 2D square lattices}

For a quantitative demonstration of the metastable electron-pair band in a crystal and its
renormalization due to many-body effects, we will use the following 2D periodic potential.
For an electron at ${\bf r}_j=(x_j, y_j)$ in a 2D square lattice with the lattice constant $\lambda$,
the considered potential is given by
\begin{equation}\label{VC}
V_c ({\bf r}_j)=V_0[\cos(q x_j)+\cos(q y_j)],
\end{equation}
where $q=2 \pi / \lambda$ and $V_0$ is the amplitude of the crystal potential.
Notice that, $V_0$ is not a measurable quantity, e.g., the amplitude of the crystal
potential in Fig.~1(c) should be infinity.
The potential defined in Eq.~(\ref{VC}) with two parameters $\lambda$ and $V_0$ will simplify
our numerical calculations without losing any essential features of the theory.
In this 2D periodic potential, the energy has a continuous
spectrum for $E \geq 0$. Therefore, two electrons can possibly bind into a pair for $E < 0$ only.
We have calculated the single-electron and metastable electron-pair states in this
periodic potential. For the calculation details we refer to Ref.~\onlinecite{hai2014}.

The single-electron states are well known in this potential.
The Schr{\"o}dinger equation for a single electron is given by
\begin{equation}
H_0\psi_{{\bf k+G}_l}({\bf r}_j)=E_l({\bf k}) \psi_{{\bf k+G}_l}({\bf r}_j),
\end{equation}
with
\begin{equation}
H_0({\bf r}_j) = - \nabla_j ^2 +V({\bf r}_j),
\end{equation}
where ${\bf k}$ is the wavevector in the first Brillouin zone, $l$ is the band index,
and ${\bf G}_l=l_x q{\bf i}+l_y q {\bf j}$ (with $l_x, l_y= 0, \pm1, \pm2,...$) is the
reciprocal-lattice vector; $E_l({\bf k})$ and $\psi_{{\bf k+G}_l}({\bf r})$ are the eigenvalue
and eigenfunction, respectively.

When we consider two electrons in this periodic potential, their Hamiltonian is given by
\begin{equation}\label{H12}
H=H_0 ({\bf r}_1)+ H_0 ({\bf r}_2) +\frac{2}{|{\bf r}_1 - {\bf r}_2 |},
\end{equation}
where the last term is the Coulomb repulsion potential between the two electrons.
In the center of mass ${\bf R}=(X,Y)$  and relative ${\bf r}=(x,y)$ coordinates defined in Eq.~(2),
the two-electron Hamiltonian becomes
\begin{eqnarray}\label{H}
\hspace*{-0.3cm}H&=&-{\frac{1}{2}} \nabla^2_{\bf R} -2 \nabla^2_{\bf r}+\frac{2}{r} \nonumber \\
&+&2V_0\left[ \cos(qX) \cos(\frac{qx}{2})+ \cos(qY) \cos(\frac{q y}{2})\right].
\end{eqnarray}
This Hamiltonian is periodic in $X$ and $Y$ with period $\lambda$. We can choose a Bloch
wavefunction in the center-of-mass coordinates for our basis. As to the function in the relative
coordinates ${\bf r}=(r,\theta)$, we have to consider the symmetry of the electron-electron Coulomb
potential and the periodic potential representing a 2D square lattice. We use
the following basis for our wavefunction,
\begin{equation}\label{psi}
\psi_{l_x,l_y;n,m}({\bf R},{\bf r})= {\frac{1}{\sqrt{A}}}e^{i({\bf k+G}_l)\cdot {\bf R}}
R_{n,m}(r) \phi_m(\theta),
\end{equation}
with
\begin{equation}\label{R}
R_{n,m}(r)=\beta c_{n,m}\left(2\beta \xi_n r \right)^m e^{-\beta \xi_n r}
L^{2m}_{n-m}(2\beta \xi_n r),
\end{equation}
and
\begin{equation}
\phi_m(\theta)=\frac{1}{\sqrt{b_m\pi}} \cos(m\theta),
\end{equation}
where $n=0,1,2,\cdots$, $m=0, 1, 2,\cdots, n$, $\xi_n=2/(2n+1)$,
$c_{n,m}=\left[2\xi_n^3 (n-m)!/(n+m)! \right]^{1/2}$, $b_0=2$, $b_m=1$ for $m\geq 1$,
and $L^{2m}_{n-m}(x)$ is the generalized Laguerre polynomial.
The function $R_{n,m}(r)$ is taken from the wavefunction of a 2D hydrogen atom\cite{2dH,Lag}
with a modification introduced by a dimensionless scaling parameter $\beta$.
The two-electron wavefunction can be written as
\begin{equation}\label{Psit}
\Psi_{\bf k}({\bf R},{\bf r})=\sum_{l_x,l_y} \sum_{n,m} a_{l_x,l_y;n,m}({\bf k})\psi_{l_x,l_y;n,m}({\bf R},{\bf r}).
\end{equation}
Considering the antisymmetry of the electron wavefunctions with spin states, the
two-electron wavefunction of the spin singlet state is given by the above expression with
the sum over even $m$ only.

Solving the corresponding eigenvalue equation of the two-electron Hamiltonian given by Eq.~(\ref{H})
with the above basis, we find a metastable electron-pair state of spin-singlet in the 2D square
lattice potential. As shown in Ref.~[\onlinecite{hai2014}], for fixed period $\lambda$,
a metastable electron-pair state can be found when $V_0$ is larger than a certain value.
A minimum potential $V_0$ required for a metastable electron-pair state in the periodic potential
corresponds to the critical nuclear charge $Z_c$ for a helium-like two-electron atom.
The metastable electron-pair state exist for $E<0$ only. This indicates that co-tunneling of
the paired electrons occurs in the formation of the electron-pair band in this 2D periodic potential.
In the calculations we found that the average separation $\langle r \rangle$ between two electrons
in a metastable pair state is always smaller than half the lattice constant, $\langle r \rangle < \lambda /2$.
The global minimum of the eigenenergy of the two-electron system occurs at $\langle r \rangle \to \infty$
corresponding to two non-interacting single electrons.
We also want to emphasize that the parameter $\beta$ in the wavefunction in Eq.~(\ref{R}) plays
the role of variational parameter to improve the correlation energy of the electron pair.
Because this parameter is directly related to the average distance between the two electrons in a pair,
it helps us to understand better the metastable electron-pair state. However,
the parameter $\beta$ does not determine the existence of the electron-pair state
in the periodic potential.

\begin{figure}[t!]
      {\includegraphics[scale=0.49]{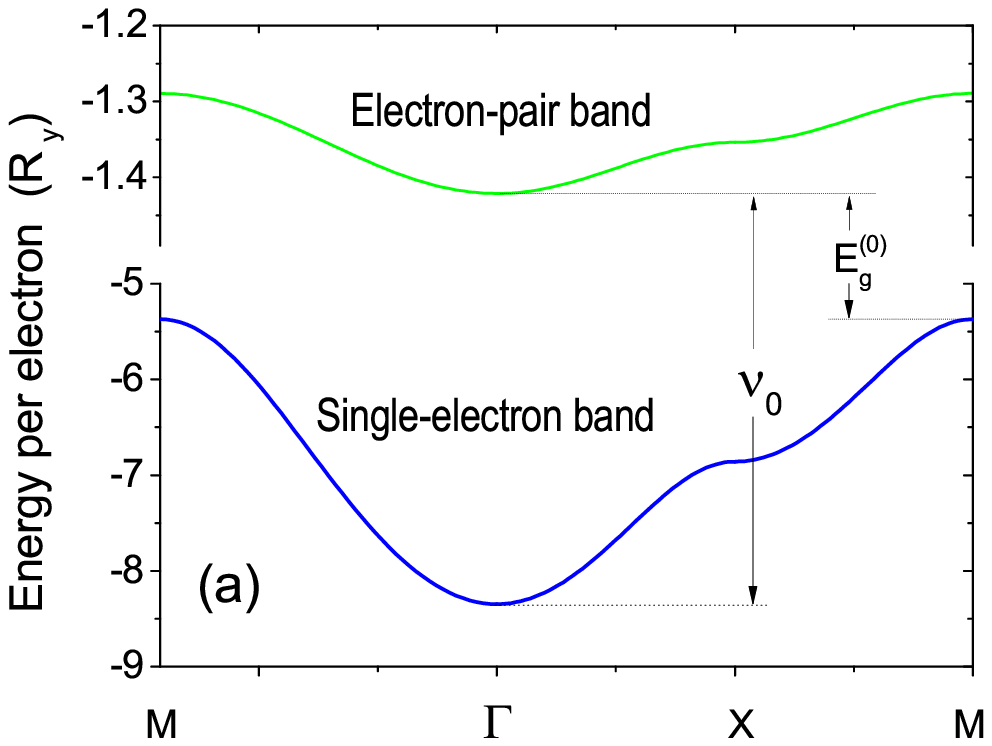}}
      {\includegraphics[scale=0.49]{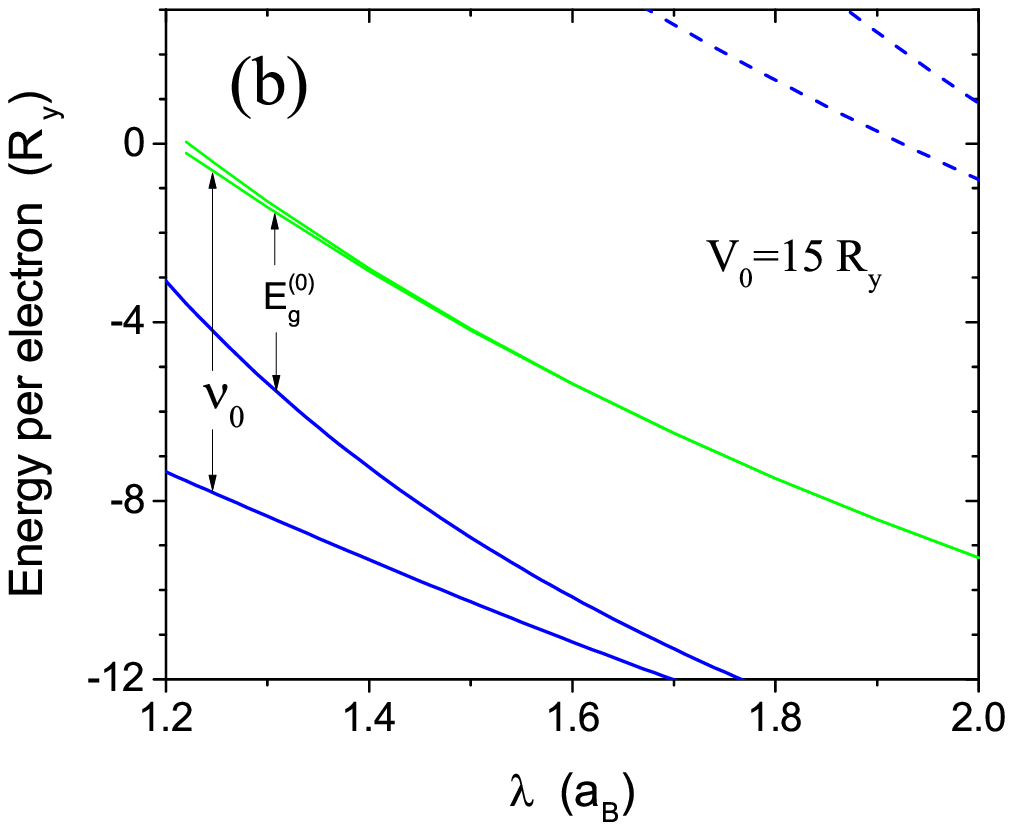}}
      {\includegraphics[scale=0.49]{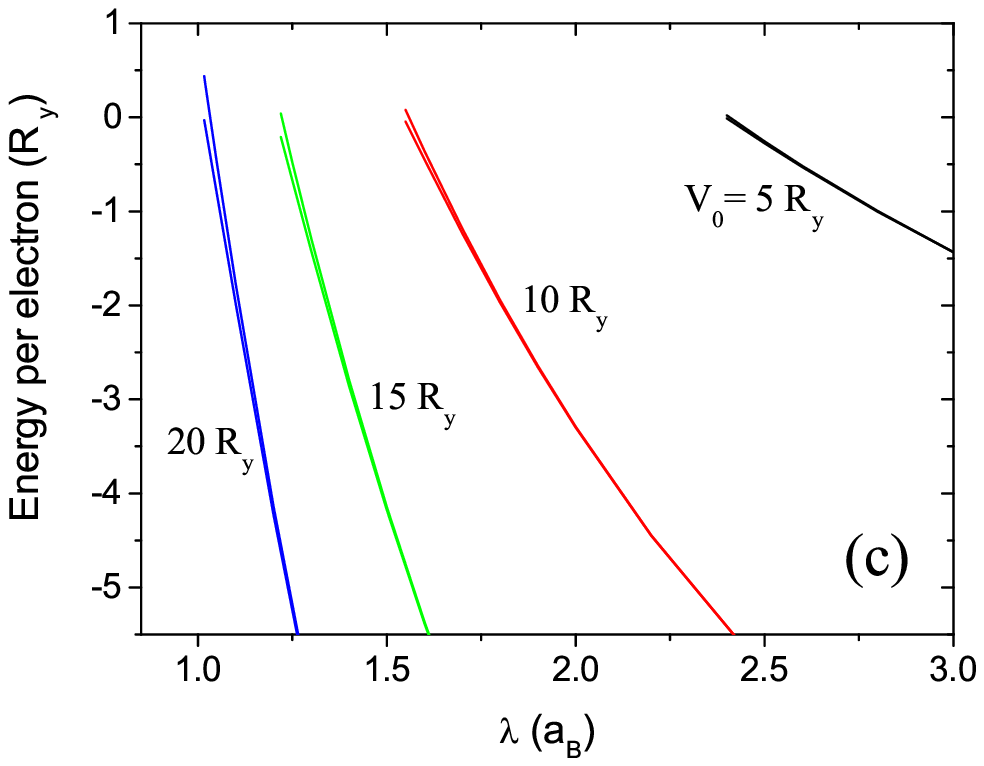}}
       {\includegraphics[scale=0.49]{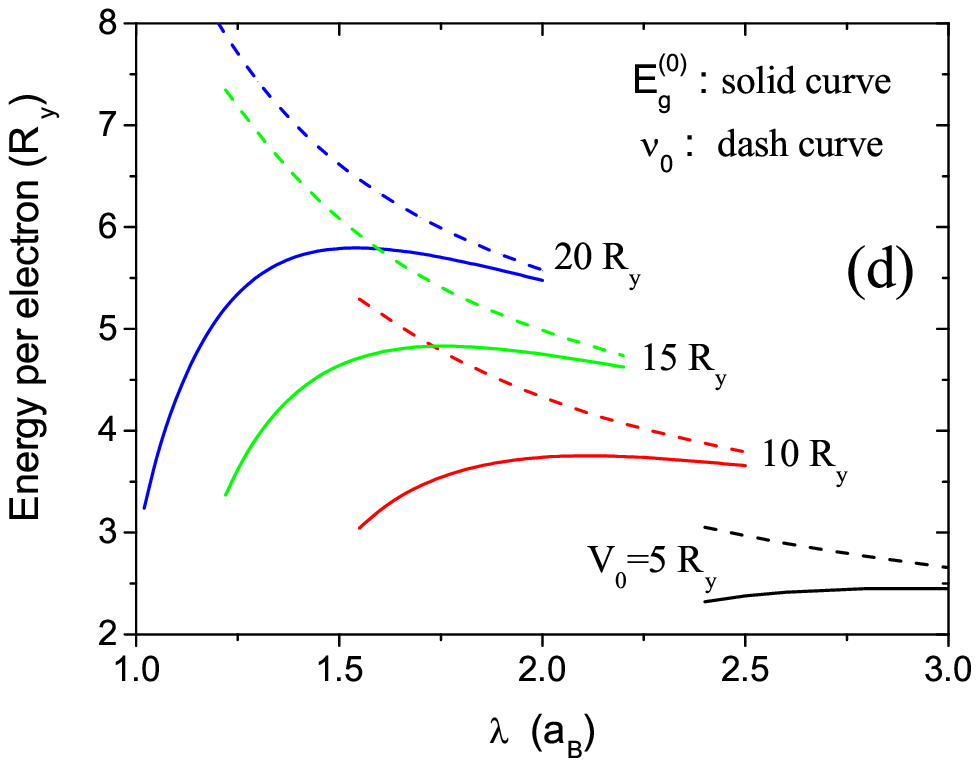}}
       \caption{(a) The dispersion relations of the electron-pair (green) and single-electron (blue)
       states in the 2D crystal with $\lambda=1.3$ a$_B$ and $V_0=15$ R$_y$.
       (b) The electron-pair band (the green curves) together with
       the two lowest single-electron bands (the solid and dash blue curves)
        versus lattice constant $\lambda$ for $V_0=15$ R$_y$. (c) The electron-pair band
        versus $\lambda$ for $V_0=5$, 10, 15, and 20 R$_y$.
       (d) The energy gap $E_g^{(0)}$ (solid curves) and $\nu_0$ (the dashed curves)
        between the electron-pair and single-electron bands
        as a function of $\lambda$ for $V_0=5$, 10, 15, and 20 R$_y$.}
       \label{figx}
   \end{figure}

In the following, we will present numerical results for the band structure and show that
the dispersion relation of the electron-pair band in this potential fits very well
the expression given by Eq.~(\ref{E2DSQ}).
Fig.~\ref{figx}(a) shows the dispersion relations of the electron states in the 2D crystal potential
with $\lambda=1.3$ a$_B$ and $V_0=15$ R$_y$. The dispersion relation $E_{p}({\bf k})$ of the spin-singlet
metastable electron-pair band is given together with that of the lowest
single-electron band $E_{0}({\bf k})$. The electron-pair band remains above
the corresponding single-electron band because the Coulomb repulsion between
the two electrons is stronger than their correlation. However, the shape of the dispersion relations of
the two bands are very similar. This confirms the dispersion relation of the electron-pair band
discussed in previous section within the framework of the tight-binding approach.
The similarity is due to the fact that two paired electrons are
closely bound in the relative coordinates in real space with an average separation $\langle r \rangle$
less than half of the lattice period. Furthermore, the single-electron and electron-pair states in
the relative coordinates are of the same symmetry. The average distance between
the two electrons in the case of Fig.~\ref{figx}(a) is $\langle r \rangle =0.44\lambda$.
The energy gap between the electron-pair band and the single-electron band is $E_g^{(0)}$.
The energy difference between the bottoms of the two bands at $\Gamma$ point is defined as $\nu_0$.
The electron pair behaves as a larger particle with both the mass and charge
twice of a single electron. Consequently, the tunneling probability
of an electron pair to its neighbor site is much smaller than that of a single electron leading to a
much narrower electron-pair band.

We also find that the dispersion relation of the electron-pair band can indeed be described by
Eq.~(\ref{E2DSQ}). For instance, the dispersion of the electron-pair band in Fig.~\ref{figx}(a)
is fitted very well by $E_{p}({\bf k})= E_{p,0}-J_{p,1}[ \cos(k_x \lambda) + \cos (k_y \lambda)]$, where
$E_{p,0}=-1.35401\pm 0.00002$ R$_y$ and $J_{p,1}=0.03368\pm 0.00002$ R$_y$. The fitting gives
an extremely small fitting parameter $\chi^2 =3\times 10^{-8}$.

In Fig.~\ref{figx}(b) we plot the electron-pair band as a function of the lattice constant $\lambda$
together with the two lowest single-electron bands for fixed potential amplitude $V_0$=15 R$_y$.
The two curves for each band indicate the minimum and maximum energies of the band.
We see that the electron-pair band appears for $\lambda \gtrsim 1.22$ a$_B$ with a bandwidth
$\lesssim 0.2$ R$_y$. It stays above the lowest single-electron band with a gap $E_g^{(0)}$ of about
3 to 5 R$_y$. The bandwidth of the electron-pair band is at least one order of magnitude smaller
than that of the single-electron band.
Fig.~\ref{figx}(c) shows the electron-pair bands as a function of $\lambda$ for different $V_0$.
We see that the electron-pair band appears for $E<0$ because the energy spectrum is continuous
for positive energy in the crystal potential given by Eq.~(\ref{VC}). It means that
co-tunneling of the paired electrons is required to form the energy band.
Therefore, their bandwidth is much smaller than that of the single-electron band.
The existence of the metastable electron-pair band is a result of the local confinement in each unit cell,
the electron-electron correlation, and the co-tunneling of the electron pair in the crystal.
In Fig.~\ref{figx}(d) we plot the energy gap $E_g^{(0)}$ together with $\nu_0$.
They are important quantities for the renormalization of the electron-pair states.

In the rest of the paper, we will demonstrate that the metastable electron pairs can be stabilized
at certain electron densities by including electron-electron and electron-phonon interactions in the crystal.
Since the electron pairs are spin singlet, they will be considered as bosonic quasiparticles
and mostly distributed at the bottom of the electron-pair band at low temperature.
The many-body effects in the crystal renormalize the band structure. If the band
renormalization can bring down the bottom of the electron-pair band to the Fermi surface of
the single-electron band, the electron pairs at the bottom of the band cannot decay into
two single electrons because of the Pauli exclusion principle.
In such a case, the electron pairs can be stabilized.

\section{Hamiltonian of the two-band many-electron system interacting with LO-phonons}

We consider the 2D electron system with two energy bands: a lower single-electron band $E_s({\bf k})$
and a higher metastable electron-pair band $\nu_0+E_p({\bf k})$.
The bottom of the single-electron band is taken as reference for energy $E=0$ and
the bottom of the electron-pair band is at $E=\nu_0$ (see Fig.~2).
Assuming that there are $N_t$ electrons in the system consisting of $N_s$ single electrons
and $N_p$ electron pairs with $N_{t}=N_s+2 N_p$, the many-particle exchange-correlation (xc) interactions
will renormalize the energy bands reducing the energies of both the single electrons and electron pairs.
In real materials, a 2D electron system can be found in the interface and surface of bulk materials or
in a 2D layer of layered crystal structure such as superconducting cuprate, therefore interaction
between the electron system and crystal lattice vibration and polarization affects the electron states.
Considering the ionic and polar-covalent characteristics of many superconducting materials,
the electron-LO-phonon interactions can be significant and their contribution
to the energy-band renormalization is important.\cite{Mahan,Devreese}
In this section, we will present the many-particle Hamiltonian including electron-electron (e-e) and
electron-phonon (e-ph) interactions assuming that the 2D electron layer is immersed in a 3D phonon field.
The Hamiltonian of the considered system is giving by
\begin{quote}
\mathchardef\mhyphen="2D
\begin{equation}\label{Htot}
{H} = {H}_{\rm el}+H_{\rm ph}+{H}_{\rm el\mhyphen ph},
\end{equation}
\end{quote}
where ${H}_{\rm el}$ is for the electronic part with single electrons and electron pairs,
$H_{\rm ph}$ for 3D LO-phonons, and ${H}_{\rm el-ph}$ for e-ph interaction.
The Hamiltonian of the electronic part ${H}_{\rm el}$ was derived in Ref.~[\onlinecite{hai2015}],
and is given by
\begin{quote}
\mathchardef\mhyphen="2D
\begin{equation}\label{Hel}
{H_{\rm el}} = {H}_{\rm single}+{H}_{\rm pair}+{H}_{\rm s\mhyphen p},
\end{equation}
\end{quote}
where $H_{\rm single}$ is for the single-electron (se) band, ${H}_{\rm pair}$ for the electron-pair (ep) band,
${H}_{\rm s-p}$ for single-electron-electron-pair (se-ep) interaction. For the $N_s$ electrons in
the single-electron band, their Hamiltonian is given by
\begin{eqnarray}\label{Hsingle}
&& {H}_{\rm single} =
\sum_{{\bf k},\sigma} E_s ({\bf k}) c^{\dag}_{{\bf k},\sigma} c_{{\bf k},\sigma} \nonumber \\
&& + \frac{1}{2A} \sum_{\bf k_1,k_2, q} \sum_{\sigma, \sigma^\prime} v_{q}
c^{\dag}_{\bf k_1 -q,\sigma} c^{\dag}_{\bf k_2 +q,\sigma^\prime}
c_{\bf k_2,\sigma^\prime}c_{\bf k_1,\sigma} \; ,
\end{eqnarray}
where $v_q =v_{ss}(q)= 2 \frac{2\pi}{q}$ is the single-electron-single-electron (se-se) Coulomb potential,
the operators $c_{{\bf k},\sigma}^\dag$ and $c_{{\bf k},\sigma}$ are
creation and annihilation operators, respectively, for a single electron of momentum $\hbar${\bf k}
and spin ${\sigma}$.
They obey the fermion anti-commutation relations
$\{ c_{{\bf k},\sigma}, c^{\dag}_{{\bf k}^{\prime},{\sigma}^{\prime} } \} =
\delta_{{\bf k},{\bf k}^{\prime}} \delta_{{\sigma},{\sigma}^{\prime}} $,
$\{ c_{{\bf k},\sigma}, c_{{\bf k}^{\prime},{\sigma}^{\prime} } \} = 0$, and
$\{ c^\dag_{{\bf k},\sigma}, c^{\dag}_{{\bf k}^{\prime},{\sigma}^{\prime} } \} = 0$.
For $N_p$ electron pairs in the electron-pair band, the Hamiltonian is given by
\begin{eqnarray}\label{Hpair}
{H}_{\rm pair} &=& \sum_{\bf k} 2 \left( \nu_0 + E_p({\bf k}) \right) b^{\dag}_{\bf k} b_{\bf k} \nonumber \\
 &+& \frac{1}{2A}\sum_{\bf k_1,k_2, q} v_{pp}(q)
b^{\dag}_{\bf k_1 -q} b^{\dag}_{\bf k_2 +q} b_{\bf k_2}b_{\bf k_1} ,
\end{eqnarray}
where $v_{pp}(q)=4 v_q f_{pp}(q)$ is the electron-pair-electron-pair (ep-ep) interaction potential
with the form factor $f_{pp}(q)$ given in Ref.~[\onlinecite{hai2015}],
the operators $b^\dag_{\bf k}$ and $b_{\bf k}$ are creation and
annihilation operators, respectively, for a spin-singlet electron pair of momentum $\hbar${\bf k}.
They obey the boson commutation relations
$ [ b_{\bf k}, b^{\dag}_{{\bf k}^{\prime}} ] = \delta_{{\bf k},{\bf k}^{\prime}}  $,
$ [ b_{\bf k}, b_{{\bf k}^{\prime}} ] =0  $, and
$ [ b^\dag_{\bf k}, b^{\dag}_{{\bf k}^{\prime}} ] = 0$.

The se-ep interband interaction is give by
\begin{quote}
\mathchardef\mhyphen="2D
\begin{equation}\label{int}
{H}_{\rm s\mhyphen p} =  {H}^{\rm s}_{\rm int} +{H}^{\rm t}_{\rm int},
\end{equation}
\end{quote}
with
\begin{eqnarray}\label{Hints}
{H}^{\rm s}_{\rm int}= \frac{1}{A}\sum_{\bf k, k_1, q,\sigma} v_{sp}(q)
b^{\dag}_{\bf k_1 -q} c^{\dag}_{\bf k +q,\sigma} b_{\bf k_1}c_{\bf k,\sigma}
\end{eqnarray}
and
\begin{eqnarray}\label{Hintt}
{H}^{\rm t}_{\rm int} &&=\frac{1}{\sqrt{A}}\sum_{\bf k, q} v^{\rm t}(q) \nonumber\\
 &&\times \left( b^{\dag}_{\bf k } c_{\frac{\bf k}{2}+{\bf q},\uparrow} c_{\frac{\bf k}{2}-{\bf q},\downarrow}
+ b_{\bf k } c^{\dag}_{\frac{\bf k}{2}-{\bf q},\downarrow}
c^{\dag}_{\frac{\bf k}{2}+{\bf q},\uparrow}  \right),
\end{eqnarray}
where $v_{sp}(q)=  2 v_q f_{sp}(q)$ is the se-ep interband scattering potential (without breaking
the electron pair) with form factor $f_{sp}(q)$ and $v^{\rm t}(q)$ is the se-ep interaction potential
for interband transition (breaking or forming electron pairs). They are given in Ref.~\onlinecite{hai2015}
and $f^2_{sp}(q)=f_{pp}(q)$.
Notice that in the above Hamiltonian, the summation over {\bf q} does not include {\bf q}=0 because it is
cancelled out with the background ion-ion interaction and the system is neutral.

As we discussed in the previous section, the single electrons and the paired electrons share
the same space in the crystal. The maximum total electron density is two electrons per unit cell (per atom),
paired or not. However, when the potential amplitude $V_0$ of the crystal is larger than a
certain value, two electrons in the same unit cell will occupy the same atomic orbital
in the relative coordinates forming an electron pair. It means that in this case, each unit
cell can be occupied by a single electron or by an electron pair, but not both at the same time.
Therefore, for a certain single-electron density $n_s=N_s/A$ (where $A$ is the area of the sample),
the maximum electron-pair density $n_p=N_p/A$ in the square-lattice crystal is given by
\begin{equation}\label{npmax}
n^{\rm max}_{p} ={\lambda^{-2} - n_s},
\end{equation}
where $\lambda^{-2}$ is the density of the unit cell of the crystal. Notice that $n_s = \lambda^{-2}$
corresponds to half filling of the single-electron band.
The above condition indicates that the single-electron band should be less than half filled
if there are any electron pairs in the crystal.

The Hamiltonian of the optical-phonon modes in bulk materials with energy $\hbar\omega_{_{LO}}$
and 3D wavevector {\bf Q} =(${\bf q},q_z$) is given by
\begin{equation}\label{Hph}
H_{\rm ph}= \sum_{\bf Q} \hbar\omega_{_{LO}} a^{\dag}_{\bf Q }a_{\bf Q } ,
\end{equation}
where $a^{\dag}_{\bf Q }$ ($a_{\bf Q } $) is the creation (annihilation) operator of the LO-phonons.
The interaction Hamiltonian of a many-electron system with the LO-phonons is given by,\cite{Mahan}
\begin{quote}
\mathchardef\mhyphen="2D
\begin{equation}\label{Heph}
H_{\rm el\mhyphen ph}= \sum_{j=1}^{N_t} \sum_{\bf Q}
\left(V_{\bf Q} a_{\bf Q } e^{i{\bf Q}\cdot{\bf r}_j} +
V^{*}_{\bf Q} a^{\dag}_{\bf Q } e^{-i{\bf Q}\cdot{\bf r}_j} \right),
\end{equation}
\end{quote}
with the Fourier coefficient of the e-ph interaction potential
\begin{equation}\label{Vph}
V_{\bf Q} = -i\hbar\omega_{_{LO}}
\left( \frac{\hbar}{2m^*\omega_{_{LO}}}\right)^{1/4} \sqrt{\frac{4\pi\alpha}{V Q^2}},
\end{equation}
where ${\bf r}_j$ is the position of
the electron $j$ with band mass $m^*$. The Fr\"ohlich electron-phonon coupling constant $\alpha$ is defined by
\begin{equation}\label{alpha}
\alpha=\frac{e^2}{\hbar}\sqrt{\frac{\hbar}{2m^*\omega_{_{LO}}}}
\left(\frac{1}{\epsilon_\infty}-\frac{1}{\epsilon_0} \right),
\end{equation}
where $\epsilon_0$ and $\epsilon_\infty$ are the static and high-frequency dielectric constant,
respectively.

In the considered two-band system, the gap between the two bands is $E^{(0)}_g$ in the zero density limit.
Although the e-e and e-ph interactions in the crystal reduce the energies of both the single-electron
and electron-pair bands, their effects on the electrons in the paired states are larger than those
in the unpaired single-electron band. Therefore, we expect that many-body effects will reduce
the energy gap between the two bands. We will calculate the exchange-correlation corrections as well as
the polaron energies for both single electrons and electron pairs including the screening effects
in order to find out whether the electron pairs can be stabilized or not.

In order to obtain the ground-state energy of the present many-particle system with the single electrons,
electron pairs and LO-phonons, we will employ the Lee-Low-Pines transformation\cite{LLP} in dealing with
the many-polaron system\cite{LLP,Lemmens,Wu,hai93}. For weak e-ph coupling, we can assume the ground
state of the electron-phonon system as $|{\rm GS} \rangle = |{\rm GS_{el}}\rangle |{\rm VAC_{ph}} \rangle $,
where $|{\rm GS_{el}}\rangle $ is the ground state of the electronic part and $|{\rm VAC_{ph}} \rangle $ is
the phonon vacuum state with zero real phonons.\cite{LLP,Lemmens} The above approximation is valid for
weak and intermediate e-ph coupling strength\cite{Devreese,LLP,Lemmens,Larsen69,Mishchenko}
and it allows us to calculate separately the electron exchange-correlation and polaron contributions
to the ground-state energy of the system, given by
\begin{equation}\label{EGS}
E_{\rm GS}= E^{\rm (el)}_{\rm GS} + E^{\rm (tot)}_{\rm pol},
\end{equation}
where $E^{\rm (el)}_{\rm GS}$ is the ground-state energy of the electronic part without interaction with
phonons and $E^{\rm (tot)}_{\rm pol}$ is the total polaron correction due to electron-phonon interaction.

In the next two sections, we will calculate the contributions of the electronic exchange-correlation
interaction and electron-phonon interaction to the renormalization of the single-electron end electron-pair
energies.

\section{Electron-electron interactions with single electrons and electron pairs}

It is known that in both 3D and 2D systems the exchange-correlation energy for band-gap renormalization
(BGR) is almost independent of the band characteristics. The many-particle exchange-correlation energy
depends only on the inter-particle distance $r_s$ (determined by the particle density)
in appropriate rescaled natural units in a universal manner\cite{VK,GT,Klein}.
The contribution of the electron-electron interaction to the BGR can be
obtained by calculating the average exchange-correlation energies per particle or by calculating the
self-energies of the particles involved. The kinetic energy
is usually assumed to be unchanged in a renormalization process.

In this section, we will calculate the ground-state energy $E^{\rm (el)}_{\rm GS}$
of the electronic part consisting of the single electrons and electron pairs.
The corresponding Hamiltonian is given by Eqs.~(24-29). Although the electron pairs are
metastable, we will treat them in the calculations as if they were stable particles.
Only the final results including full many-body corrections will tell us whether they
can really be stabilized or not.
In the calculations, we will first take the single-electron density $n_{s}$ and
electron-pair density $n_p$ as inputs. The single electrons are considered as fermions
and electron pairs as bosons. Therefore, we are dealing with a many-particle system
consisting of a boson-fermion mixture.\cite{Friedberg}
The exchange-correlation energies are obtained as a function of $n_{s}$ and $n_p$. We then
determine their contributions to the ground-state energy and the band renormalization.
Within such a scheme, the se-ep interaction Hamiltonian $H^{t}_{\rm int}$ given in Eq.~(\ref{Hintt})
will not be invoked explicitly in the calculations. However, transitions between the single-electron
and electron-pair bands are permitted because of this term.
˜
The many-particle interaction energy in such a two-component system of boson-fermion mixture
can be obtained by\cite{pines1966}
\begin{equation}\label{EPines}
E_{ij}= \int_0^{e^2} \frac{E^{\rm int}_{ij} ( \xi )}{\xi} d\xi,
\end{equation}
where the inter-particle interaction potential $E^{\rm int}_{ij} (\xi)$ is given by
\begin{eqnarray}
E^{\rm int}_{ij}(e^2) = \frac{1}{2A} \sum_{\bf q} v_{ij}(q)[S_{ij}(q)-\delta_{ij}] .
\end{eqnarray}
The above potential depends on the ``bare'' inter-particle potential $v_{ij}(q)$ and
static structure factor $S_{ij}(q)$, for $i, j =s$ (single electron) and
$p$ (electron pair). The static structure factor can be calculated by,
\begin{equation}\label{SSF}
S_{ij}(q)= -\frac{1}{\pi\sqrt{n_i n_j}} \int_0^{\infty} d\omega \chi_{ij}(q, i\omega).
\end{equation}
Within the linear response theory, the density-density response function $\chi_{ij}(q,\omega)$
of this two-component plasma is given by\cite{PV,AT},
\begin{equation}\label{chi}
\{[\chi(q,\omega)]^{^{-1}}\}_{ij}=[\chi^{(0)}_{ii}(q,\omega)]^{^{-1}}\delta_{ij} -\varphi_{ij}(q),
\end{equation}
where $\chi^{(0)}_{ii}(q,\omega)$ is the non-interacting polarizability\cite{PV,AT,Ando82,Hines}
of the $i$th component and $\varphi_{ij}(q)$ is the static effective interaction potential.
The function $\chi^{(0)}_{ss}(q,\omega)$ is for a non-interacting 2D electron gas given in Ref.~\cite{Ando82}.
The function $\chi^{(0)}_{pp}(q,\omega)$ is the polarizability of a non-interacting 2D boson gas
of electron pairs with density $n_p$.
The effective potential $\varphi_{ij}(q)$ defines a local field correction\cite{PV,Moudgil} in terms of
the ``bare'' potential $v_{ij}(q)$. Within the random-phase approximation (RPA),
the local field correction on the static effective interaction is neglected and therefore
$\varphi_{ij}(q)=v_{ij}(q)$. The potential $v_{ij}(q)$ has been determined in the previous section
given by $v_{ss}(q)=v_q$, $v_{pp}(q)=4 v_q f_{pp}(q)$, and $v_{sp}(q)=v_{ps}(q)=2 v_q f_{sp}(q)$.

For a non-interacting 2D boson (i.e., electron-pair) gas with density $n_p$,
we can assume that all the electron pairs are in the same state at the bottom of
the electron-pair band at zero temperature, i.e., in the condensate phase\cite{AT,Hines}.
The polarizability of the non-interacting boson gas is given by,
\begin{equation}\label{chi0biw}
\chi^{(0)}_{pp}(q,i\omega) = -\frac{2 n_p \varepsilon_{q,p}}{\omega^2+\varepsilon^{2}_{q,p}},
\end{equation}
where $\varepsilon_{q,p}=q^2 /2 $.

The contribution of the e-e interaction to the single-electron band renormalization
is given by $\Delta E_s = E_{ss}+E_{sp}$ and to the electron-pair band given by $\Delta E_p = (E_{pp}+E_{sp})/2$.
The ground-state energy of the two-band system is given by
\begin{equation}\label{EGSel}
E^{\rm (el)}_{\rm GS} = E_{\rm kin}+N_s(E_{ss}+E_{sp})+2 N_p\left(\nu_0+\frac{E_{pp}+E_{sp}}{2}\right),
\end{equation}
where $E_{\rm kin}$ is the kinetic energy of the many-electron system.
We calculate these energies within the RPA. Although the RPA overestimates the exchange-correlation
energies $\Delta E_s$ and $\Delta E_p$, only the difference between them contributes to
the band-gap renormalization. The errors resulting from the RPA should be partially cancelled in the
process when determining the condition of stability of the electron pairs.
Therefore, we consider the RPA a reasonable approximation for our purpose.
As mentioned above, the kinetic energy $E_{\rm kin}$ will be assumed unchanged in the renormalization.
It is given by $E_{\rm kin}= N_s \varepsilon_F /2$, where $\varepsilon_F$ is the Fermi energy of
the single-electron band in relation to its band bottom. In a two-dimensional system, the average kinetic
energy of a single electron is $\varepsilon_F /2$. The electron pairs have no kinetic energy
because they are assumed to be at the bottom of the electron-pair band in the condensate phase.

\begin{figure}[b!]
      \includegraphics[scale=0.7]{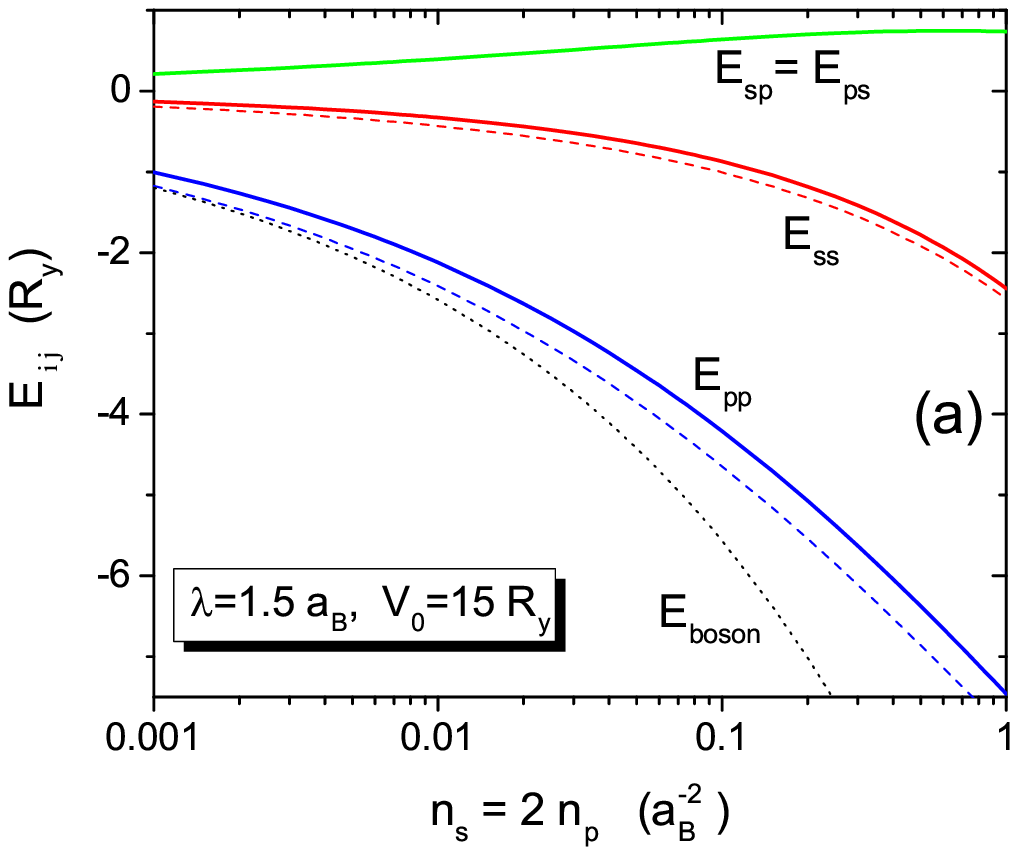}
      \includegraphics[scale=0.7]{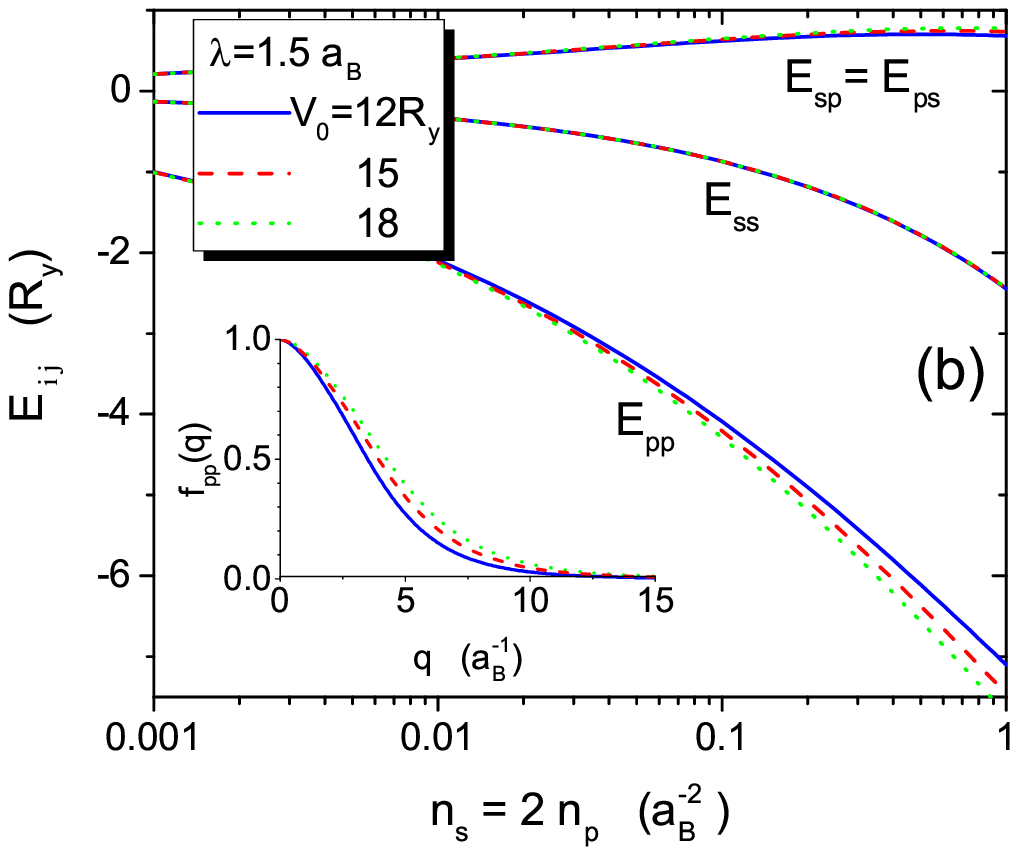}
       \caption{(a) The many-particle interaction energies $E_{ss}$, $E_{sp}$, $E_{ps}$, and $E_{pp}$
       in the 2D square-lattice potential with $\lambda=1.5$ a$_B$ and $V_0=15$ R$_y$ keeping $n_s=2 n_p$.
       The red-dash (blue-dash) curve is for $E_{ss}$ ($E_{pp}$) without se-ep interaction.
       The black-dotted curve is for the 2D ideal charged boson gas.
       (b) The interaction energies in the 2D systems with $\lambda=1.5$ a$_B$ and different $V_0$.
       The form factor $f_{pp}(q)$ is given in the inset.}
       \label{IntEGS}
   \end{figure}

If we further assume that $v_{sp}(q)=0$, the single electrons and electron pairs become
independent in different energy bands. The single-electron system is an usual one-component
2D electron gas which has been widely investigated within different methods and
approximations.\cite{Ando82,isihara,ceperley}
The system of electron pairs is new. Although the electron pairs are metastable states,
we will treat them as stable particles when searching for their many-particle ground state.
The static structure factor of a one-component boson system of electron pairs within the RPA
is given by,
\begin{equation}
S^{(1)}_{pp}(q) = [1+2 n_p v_{pp}(q) /\varepsilon_{q,p}]^{-\frac{1}{2}}.
\end{equation}
Consequently, we obtained the many-electron-pair exchange-correlation energy,
\begin{equation}\label{Epp0}
E^{(1)}_{pp} = 8(r^{p}_{s})^{^{-\frac{2}{3}}} I,
\end{equation}
where $r^p_{s}=(\pi n_p)^{-\frac{1}{2}}$, and
\begin{equation}\label{Iint}
I=\int_{0}^{\infty} dx
\{ \left(\frac{x^3}{4}\right)\left[\sqrt{1+ \frac{8}{x^3} f_{pp}(q)} -1 \right]-f_{pp}(q) \}.
\end{equation}
The calculations in Ref.~[\onlinecite{hai2015}] showed that $f_{pp}(q)$ is a monotone decreasing function
with $f_{pp}(0)=1$ and $f_{pp}(\infty)=0$. If we take $f_{pp}(q)\equiv 1$ (consequently $\langle r \rangle= 0$),
i.e., assuming the electron pair as an ideal boson with charge -2$e$ and mass 2$m^*$,
the above integral becomes $ I=I_0 = -1.29 $.
This is the well-known RPA result for an ideal charged 2D boson gas\cite{Um}. Notice that the
factor 8 in Eq.~(\ref{Epp0}) is due to the units used here.

In Fig.~\ref{IntEGS} we show the many-particle interaction energies $E_{ss}$, $E_{sp}$, and $E_{pp}$
within the RPA in the system keeping the same number of electrons in the single-electron and electron-pair bands,
i.e., $n_s = 2 n_p$. The energy $E_{sp}=E_{ps}$ is positive because of the se-ep repulsion.
Fig.~\ref{IntEGS}(a) is for the 2D crystal with $\lambda=1.5$ a$_B$ and $V_0=15$ R$_y$.
It shows the density dependence of the interaction energies and the effects of se-ep interaction.
The blue-dash and red-dash curves are the exchange-correlation energies $E^{(1)}_{pp}$ and
$E^{(1)}_{ss}$, respectively, without the se-ep interaction.
In this case, the energy $E^{(1)}_{pp}$ is obtained from Eqs.~(\ref{Epp0}) and (\ref{Iint}).
We see that the nonzero distance between the two electrons in the pair
(i.e., $\langle r \rangle >0$) affects the energy $E_{pp}$. In the calculations
we found that $\lambda /2 > \langle r \rangle > \lambda /3$.
If we assume $\langle r \rangle = 0$ for the electron pairs, we obtain
$E^{(1)}_{pp} =E_{\rm boson}= -1.29\times 8(r^{p}_{s})^{^{-\frac{2}{3}}} $
for an ideal charged boson system in 2D indicated by the black-dotted curve.

We observe that the se-ep interband interaction not only introduces the energy $E_{sp}$ but also
reduces the energies $E_{pp}$ and $E_{ss}$ being evident in the difference between the solid and dashed
curves in Fig.~\ref{IntEGS}(a). The density dependence of the ep-ep interaction energy $E_{pp}$
is different from that of the se-se interaction $E_{ss}$.
For example, $E_{pp}$ is about 7 times larger than $E_{ss}$ at lower density $n_s = 2 n_p=0.001$ a$^{-2}_B$
and this ratio is reduced to 3 times at higher density $n_s = 2 n_p=1.0 $ a$^{-2}_B$.
In Fig.~\ref{IntEGS}(b) we show the many-particle interaction energies for different potential $V_0$.
With increasing $V_0$ for the same lattice constant $\lambda$, the average distance $\langle r \rangle$
between the two electrons in the same pair decreases and, consequently,
the form factor $f_{pp}(q)$ increases (as shown in the inset) and the energy $E_{pp}$ becomes larger.

\section{Polaron effects in the two-band system with single electrons and electron pairs}

In this section, we will study the polaron effects on the $N_t$ electrons in the two-band system
interacting with LO-phonons assuming $N_s$ electrons in the single-electron band and $N_p$
electron pairs in the electron-pair band, $N_s+2N_p=N_t$. The Hamiltonian of the system is
given in Eq.~(\ref{Htot}) and the electron-phonon interaction given by Eq.~(\ref{Heph}).
Considering a boson-fermion mixture (namely, the electron pairs and single electrons)
interacting with the phonons, the electron-phonon interaction Hamiltonian in Eq.~(\ref{Heph})
can be separated into two parts,
\begin{quote}
\mathchardef\mhyphen="2D
\begin{equation}\label{hep2p}
H_{\rm el\mhyphen ph}=  H^{\rm single}_{\rm e\mhyphen ph}+ H^{\rm pair}_{\rm e\mhyphen ph} ,
\end{equation}
\end{quote}
with
\begin{quote}
\mathchardef\mhyphen="2D
\begin{equation}\label{Hseph}
  H^{\rm single}_{\rm e\mhyphen ph}= \sum_{j=1}^{N_s} \sum_{\bf Q}
\left(V_{\bf Q} a_{\bf Q } e^{i{\bf Q}\cdot{\bf r}_j} +
V^{*}_{\bf Q} a^{\dag}_{\bf Q } e^{-i{\bf Q}\cdot{\bf r}_j} \right),
\end{equation}
\end{quote}
for $N_s$ single electrons, and
\begin{quote}
\mathchardef\mhyphen="2D
\begin{eqnarray} \label{Hepph}
&& H^{\rm pair}_{\rm e\mhyphen ph} = \sum_{j=1}^{N_p} \sum_{\bf Q} \left[\sum_{\nu=1}^{2}
\left(V_{\bf Q} a_{\bf Q } e^{i{\bf Q}\cdot{\bf r}_{j,\nu}} +
V^{*}_{\bf Q,} a^{\dag}_{\bf Q } e^{-i{\bf Q}\cdot{\bf r}_{j,\nu}} \right)\right]\nonumber\\
&&= \sum_{j=1}^{N_p} \sum_{\bf Q}
\left( V^p_{\bf Q}({\bf r} ) a_{\bf Q } e^{i{\bf Q}\cdot{\bf R}_{j}} +
V^{p^*}_{{\bf Q}}({\bf r} )  a^{\dag}_{\bf Q } e^{-i{\bf Q}\cdot{\bf R}_{j}} \right)
\end{eqnarray}
\end{quote}
for $N_p$ electron pairs, where $V^p_{\bf Q}({\bf r} )=2 V_{\bf Q} \cos(\frac{{\bf Q}\cdot{\bf r}}{2})$
and ${\bf R}_j$ and {\bf r} are the center-of-mass and relative coordinates of the electron pair, respectively.

In the calculations of the polaron energies, we will ignore the direct participation
of the single-electron-electron-pair interaction potential $v_{sp}(q)$.
The potential $v_{sp}(q)$ has a minor effect on the e-ph interaction.
Its influence on the e-ph interaction is mostly indirect through the electronic screening
and it is taken into account in the static structure-factor.
Within such an approximation - ignoring the potential $v_{sp}(q)$ in dealing with the e-ph interactions,
the Hamiltonian of the whole system defined in Eq.~(\ref{Htot}) can be separated into two subsystems.
They are the one consisting of single electrons interacting with phonons and
the other of electron pairs interacting with phonons.
In this way, the contribution of the electron-phonon interaction to
the ground-state energy $E^{\rm tot}_{\rm pol}$ in Eq.~(\ref{EGS}) can be calculated as
\begin{equation}\label{Epol}
 E^{\rm (tot)}_{\rm pol} = N_s E^{\rm single}_{\rm pol}+ N_p E^{\rm pair}_{\rm pol},
\end{equation}
where $E^{\rm single}_{\rm pol}$ and $E^{\rm pair}_{\rm pol}$ are the polaron energies of the
single electron and electron pair, respectively.

The subsystem composed of single electrons interacting with LO-phonons is
\begin{quote}
\mathchardef\mhyphen="2D
\begin{equation}\label{seph}
H^{\rm single}_{\rm pol} = {H}_{\rm single} + H_{\rm ph} + H^{\rm single}_{\rm e\mhyphen ph},
\end{equation}
\end{quote}
where $H_{\rm single}$ is given by Eq.~(\ref{Hsingle}),  $H_{\rm ph}$ by Eq.~(\ref{Hph}),
and $H^{\rm single}_{\rm e-ph}$ by Eq.~(\ref{Hseph}). The e-ph coupling leads to
a polaron consisting of an electron and a surrounding phonon cloud.
When the e-ph interaction is not too strong, the polaron correction to the ground-state energy
of an electron gas can be calculated within the Lee-Low-Pines (LLP)
unitary transformation method.\cite{LLP,Lemmens}
This method has been used to study polaron gases in bulk materials and also in low-dimensional
systems.\cite{Wu,hai93} It is known that the polaron energy obtained from the LLP transformation
is exact for $\alpha \rightarrow 0 $. In the low electron density limit, the polaron energy obtained
from the LLP method for e-ph coupling constant $\alpha=6$ is 90\% of the exact value.\cite{Devreese,Mishchenko}
Here we are dealing with a polaron gas in which the screening reduces the electron-phonon interaction
strength. Therefore, the LLP method should yield a reasonable polaron energy for $\alpha <6$.

We calculate the polaron energy within the LLP method for the above Hamiltonian in Eq.~(\ref{seph}),
given by,
\begin{equation}\label{polEs}
  E^{\rm single}_{\rm pol} = - \sum_{{\bf q},q_z} {{|V_{\bf Q}|^2 S_{ss}^2(q)}
  \over{\hbar\omega_{_{LO}}S_{ss}(q)+{\hbar^2 q^2}/{2m^*}}},
\end{equation}
where $S_{ss}(q)$ is the static structure factor of the single-electron gas.
In the low electron-density limit, $S_{ss}(q)=1$. This leads to the well known
perturbation result $ E^{\rm single}_{\rm pol}=-(\pi/2)\alpha\hbar\omega_{_{LO}}$
for the polaron energy of an electron in 2D coupled with 3D-phonons.\cite{Wu,Sarma85}

\begin{figure}
      \includegraphics[scale=0.7]{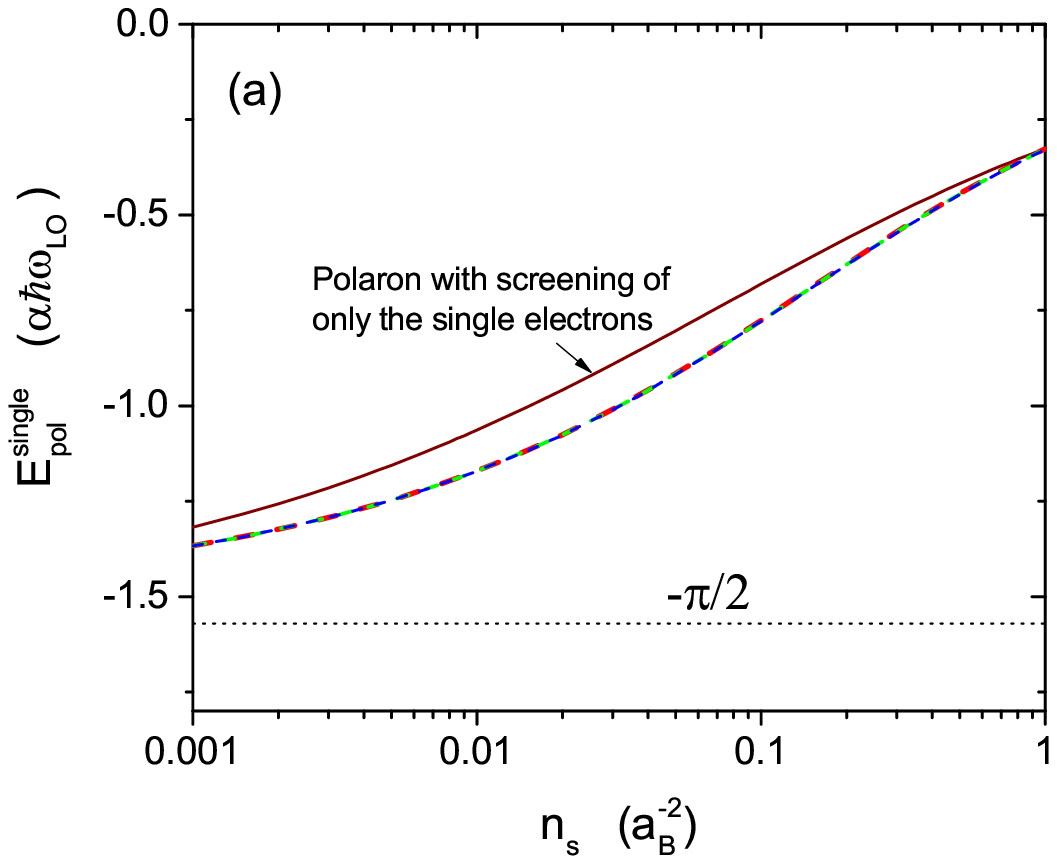}
      \includegraphics[scale=0.7]{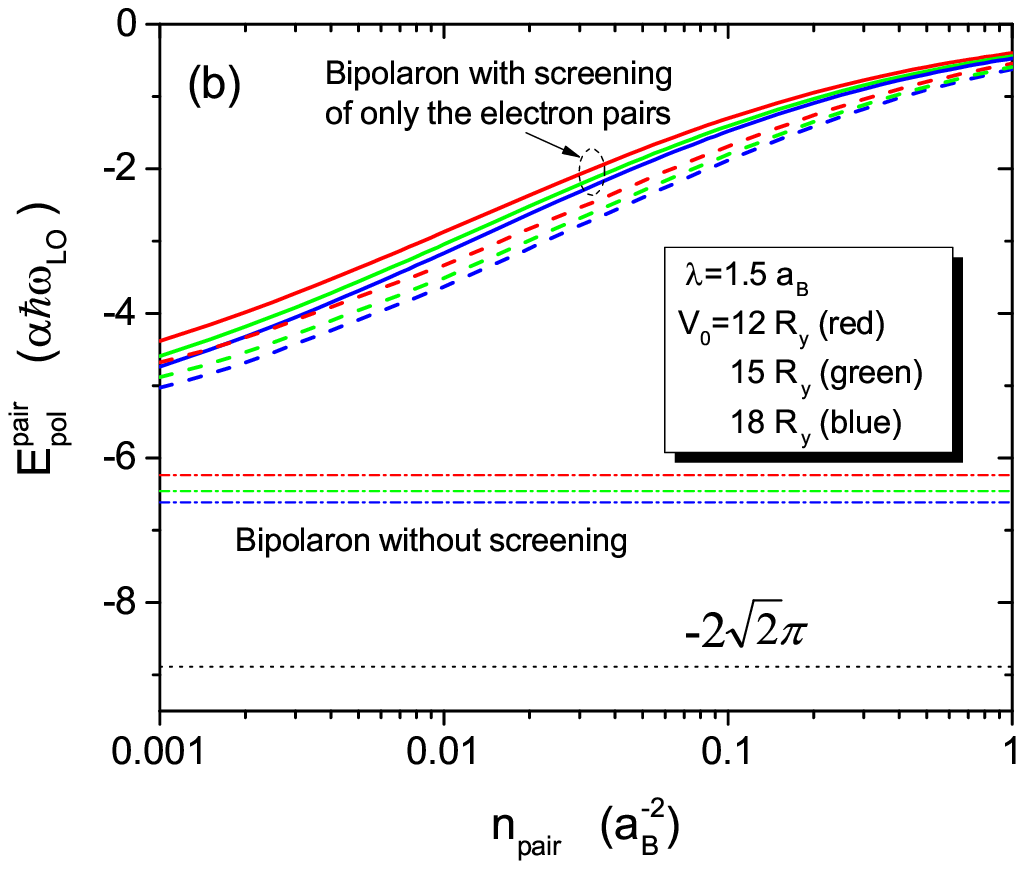}
       \caption{(a) The polaron energy as a function of the electron density $n_s$ for $\lambda=1.5$ a$_B$.
       The brown-solid curve is for polarons with screening of the single-electron gas only.
       The dashed curves are for polarons with screening of the coupled single-electrons and
       electron pairs with $n_p=n_s/2$.
       The red, green, and blue dashed curves are for $V_0=12$, 15, and 18 R$_y$, respectively.
       (b) The bipolaron energy as a function of the electron-pair density $n_p$.
       The solid (dash) curves for screening of the electron pairs without (with) interaction with single
       electrons ($n_s=2n_p$). The horizontal dash-dotted lines are for bipolarons without screening.
       The horizontal black-dotted line is the upper bound of the bipolaron energy.}
       \label{Pol}
\end{figure}

Fig.~\ref{Pol}(a) shows the polaron energy as a function of the single-electron density $n_s$ in
the 2D square-lattice crystal with $\lambda=1.5$ a$_B$ coupled with the 3D LO-phonons.
The polaron energy in the low-density limit without screening is indicated by the thin-dotted line.
It is seen that the screening of the electron gas considered in the structure factor $S_{ss}(q)$
in Eq.~(\ref{polEs}) reduces the polaron effect. At higher electron density
$n_s=1.0$ a$_B^{-2}$, the polaron energy is only about 20\% of its low-density value.
When there are also electron pairs in the crystal and the se-ep interaction is included in
the structure factor $S_{ss}(q)$ in the screening, the calculation results show that the se-ep interaction
reduces the screening and consequently, enhances the polaron energy.
The dashed curves in Fig.~\ref{Pol}(a) are obtained with electron-pair density $n_p=n_s/2$ interacting with
single-electrons. The red, green and blue dashed curves are for $V_0=12$, 15, and 18 R$_y$, respectively.
The potential $V_0$ affects the potentials $v_{sp}$ and $v_{pp}$. Its influences on the
polaron energy is indirectly through the structure factor $S_{ss}(q)$. Therefore, we obtained almost the
same value of $ E^{\rm single}_{\rm pol}$ for different $V_0$.

The subsystem consisting of electron pairs and LO-phonons is given by the
following Hamiltonian,
\begin{quote}
\mathchardef\mhyphen="2D
\begin{equation}
H^{\rm pair}_{\rm pol} =  H_{\rm pair} + H_{\rm ph}  + H^{\rm pair}_{\rm e\mhyphen ph},
\end{equation}
\end{quote}
where $H_{\rm pair}$ is given by Eq.~(\ref{Hpair}), $H_{\rm ph}$ by Eq.~(\ref{Hph}), and
$ H^{\rm pair}_{\rm e-ph}$ by Eq.~(\ref{Hepph}).
Following a similar procedure as before and applying the LLP transformation to the electron-pair-phonon
interactions, we obtain the polaron contribution to the energy of the electron pairs,
given by
\begin{equation}\label{polEp}
  E^{\rm pair}_{\rm pol} = - \sum_{{\bf q},q_z} { \left| M_{\rm pair}(\bf Q)\right|^2 S^2_{pp}(q)
  \over
  {\hbar\omega_{_{LO}}S_{pp}(q) +{\hbar^2 q^2}/{4m^*}}},
\end{equation}
where $S_{pp}(q)$ is the static structure factor of the electron pairs. The matrix element is given by
\begin{equation}
\left| M_{\rm pair}(\bf Q)\right|^2 =
| \langle  \Psi_{{\bf k}-{\bf q}}({\bf R},{\bf r});{\bf Q}
| H^{\rm pair}_{\rm e-ph} |\Psi_{\bf k}({\bf R},{\bf r}); 0 \rangle |^2,
\end{equation}
where $| \Psi_{\bf k}({\bf R},{\bf r});{\bf Q} \rangle = |\Psi_{\bf k}({\bf R},{\bf r}) \rangle |{\bf Q} \rangle $,
with $|\Psi_{\bf k}({\bf R},{\bf r}) \rangle$ for the electron-pair state and $|{\bf Q} \rangle $ for the phonon
state. Using the metastable electron-pair wavefunction in Eq.~(22), we obtain
\begin{equation}
\left| M_{\rm pair}(\bf Q)\right|^2  =4 |V_{\bf Q}|^2 f_{pp}(q) ,
\end{equation}
where $V_{\bf Q}$ is the e-ph interaction potential given by Eq.~(\ref{Vph}),
$f_{pp}(q)$ is the form-factor that appears in the pair-pair interaction
potential in Eq.~({\ref{Hpair}}).

We now obtain a Fr\"ohlich bipolaron formed by an electron pair coupled with LO-phonons.
Fig.~\ref{Pol}(b) shows the bipolaron energy in the crystals with  $\lambda=1.5$ a$_B$ and
$V_0=12$, 15, and 18 R$_y$. The solid curves
give the bipolaron energies for different $V_0$ with only screening of the electron pairs.
For larger $V_0$, the average distance $ \langle  r \rangle $ between the two electrons in a pair is smaller,
i.e., the size of the electron pair is smaller. Therefore, the bipolaron energy is larger.
But the screening reduces significantly the bipolaron energy.
We also see that, when se-ep interaction is included in the structure factor $S_{pp}(q)$, the screening
of the electron-pair-phonon coupling is reduced enhancing the bipolaron energy as shown
by the dashed curves for $n_s=2 n_p$ in the figure.

If we assume $S_{pp}(q)=1$, we obtain the low-density limit of the bipolaron energy without screening
given by the horizontal dotted-dash lines. This is the case of a single bipolaron without screening.
If we further assume $f_{pp}(q)=1$, we obtain the upper bound of the bipolaron energy,
$E^{\rm pair}_{\rm pol}  = -4\sqrt{2} (\pi/2) \alpha\hbar\omega_{_{LO}}$.\cite{Shi97}
This is equivalent to assuming the electron pair as a ``larger single electron" with charge
-2$e$ and mass 2$m^*$ coupled to the LO-phonons.

As a matter of fact, the effects of electron-phonon interaction and possible bipolaron
formation have been extensively studied as the electron pairing mechanism for unconventional
superconductivity.\cite{Devreese,Emin,Adamowski,Verbist}
In order to form a bipolaron in the crystal, a crucial point is that the electron-phonon coupling induced
attraction has to overcome the electron-electron Coulomb repulsion.\cite{Devreese}
This requires not only a small ratio $\epsilon_\infty/\epsilon_0$ of dielectric constants but also a
large enough electron-phonon coupling constant $\alpha$ leading to a critical
e-ph coupling constant $\alpha_c=2.9$ in 2D and $\alpha_c=6.8$ in 3D for bipolaron formation.

However, the bipolaron formation mechanism in the present theory is distinct from the traditional
bipolaron theory in the literature. In this paper, we show a preformed metastable electron
pair due to strong correlation of two electrons occupying the same orbital. The electron-phonon
interaction dresses the electron pairs up with a phonon cloud forming bipolarons.
Therefore, in the present context we introduce an internal interaction due to orbital dependent
electron correlation of the electron pair. The electron-phonon coupling involves in
the renormalization of the preformed electron pairs. Notice that polaron effects can be much larger
on the electrons in the correlated pairs than on the single electrons, and therefore the bipolaron
contribution to the stabilization of the electron pairs overcoming the Coulomb repulsion becomes essential.
We will show in the next section that the certain electron pairs can indeed be stabilized as bipolarons.

\section{Stabilization of the electron pairs and the ground state of the many-particle system}

The condition for stabilization of the electron pairs when including renormalization is that
the bottom of the electron-pair band occurs at the Fermi energy of the single-electron band.
Using the same energy reference defined in Sec. III, i.e., taking the bottom of the single-electron
band at $\Gamma$ point before the renormalization as $E=0$, the many-particle interactions
lower the single-electron band bottom to
\begin{equation}
   E^{\Gamma}_{\rm s} = E_{ss}+E_{sp}+E^{\rm single}_{\rm pol},
\end{equation}
where $E_{ss}$ and $E_{sp}$ are determined by Eq.~(\ref{EPines}) and $E^{\rm single}_{\rm pol}$ is given
by Eq.~(\ref{polEs}). The Fermi energy of the single-electron band can be obtained
by $E_F = E^{\Gamma}_{\rm s} +\varepsilon_F$
where $\varepsilon_F$ is the energy difference between the Fermi energy and the bottom of the band.
It is determined by the single-electron density and the density of states of the band.
We will not consider the many-body effects on $\varepsilon_F$.
The average kinetic energy per single electron in this band is given by $\varepsilon_F /2$
in a two-dimensional system.

\begin{figure}[t!]
     \includegraphics[scale=0.8]{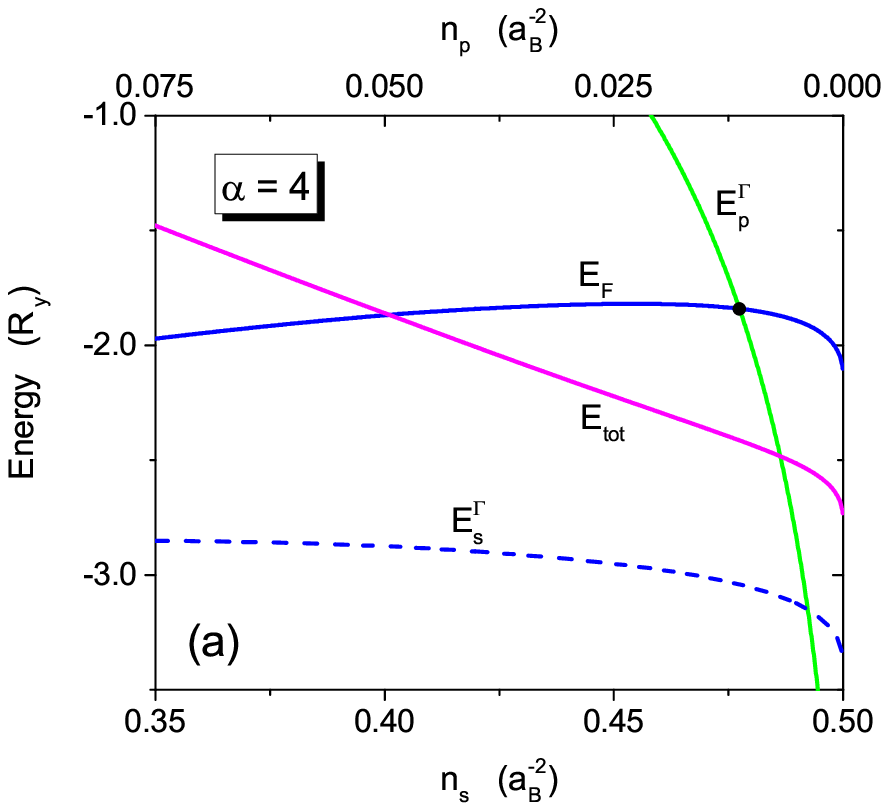}
      \includegraphics[scale=0.8]{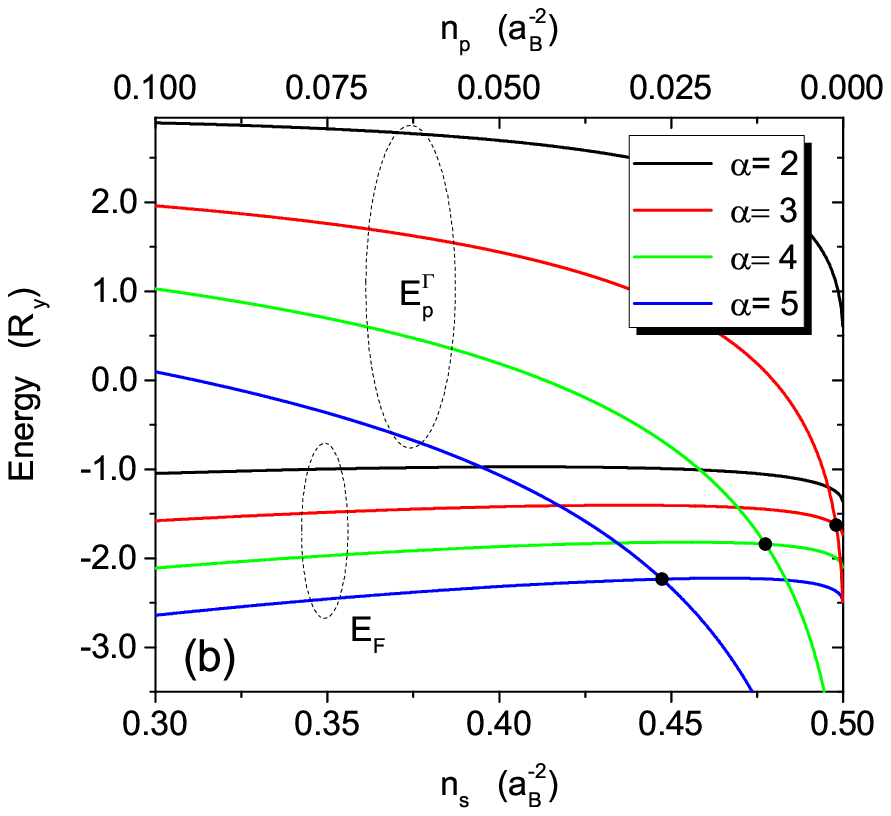}
       \caption{The renormalized energies in the system of total electron density
       $n_t = n_s+2n_p= 0.5$ a$_B^{-2}$ in the 2D crystal with $\lambda=1.3$ a$_B$ and
       $V_0=15$ R$_y$ coupled to phonons. (a) The bottom of single-electron band $E^{\Gamma}_{\rm s}$,
       the Fermi energy $E_F$, the bottom of the electron-pair band $E^{\Gamma}_{\rm p}$, and
       the total energy per electron $E_{\rm tot}$ are shown for e-ph coupling constant $\alpha=4$.
       (b) The bottom of the electron-pair band and the Fermi energy of the single-electron band for
       different values of the coupling constant $\alpha$. The black dots indicate the position where
       $E^{\Gamma}_{\rm p}=E_F$.}
       \label{Ealfa}
\end{figure}

Including the band renormalization, the bottom of the electron-pair band is given by
\begin{equation}
  E^{\Gamma}_{\rm p} = \nu_0 +\frac{_1}{^2} (E_{pp}+E_{sp}+E^{\rm pair}_{\rm pol}),
\end{equation}
where $E_{pp}$ and $E_{sp}$ are given by Eq.~(\ref{EPines}) and  $E^{\rm pair}_{\rm pol}$ by
Eq.~(\ref{polEp}). Notice that the energy of an electron pair is given by 2$E^{\Gamma}_{\rm p}$.

Fig.~{\ref{Ealfa}} shows the energies for the renormalized single-electron and electron-pair bands
in the 2D crystal with $\lambda=1.3$ a$_B$ and $V_0=15$ R$_y$ coupled with the 3D LO-phonons.
The band structure without many-body corrections was given in Fig.~2(a).
The many-particle interaction energies are obtained for a constant total electron density
$n_t = n_s+2n_p= 0.5$ a$_B^{-2}$ with different e-ph coupling constant $\alpha$.
Because the polaron energy is given by the coupling constant $\alpha$ and the LO-phonon
energy $\hbar\omega_{LO}$,  the ratio R${_y}/ \hbar\omega_{LO}$ is required in the calculations.\cite{PhLO}
This ratio is a material dependent parameter and has been used in the bound-polaron and
bound-bipolaron problems\cite{PhLO,Kashirina}. Its value is in the range from 0.5 to 2 for
different materials.\cite{Kashirina,Heumen} For the present discussion, we take R${_y}/ \hbar\omega_{LO}$=1.

\begin{figure}[hbt!]
      \includegraphics[scale=0.67]{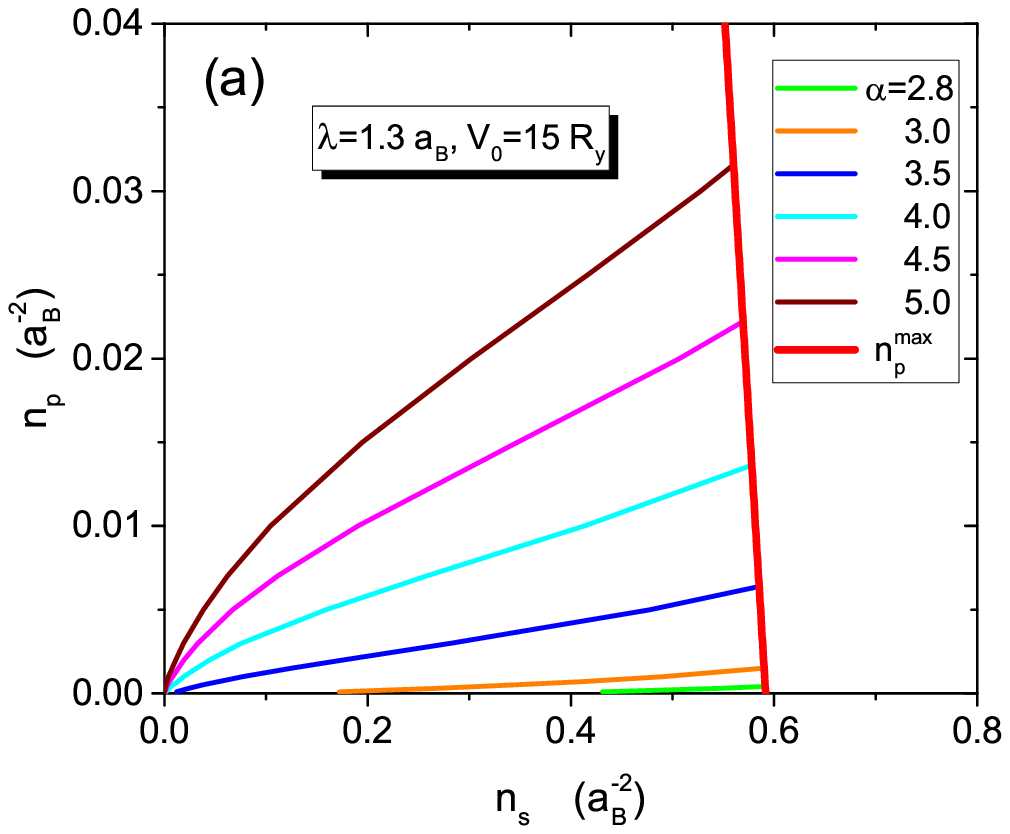}
      \includegraphics[scale=0.67]{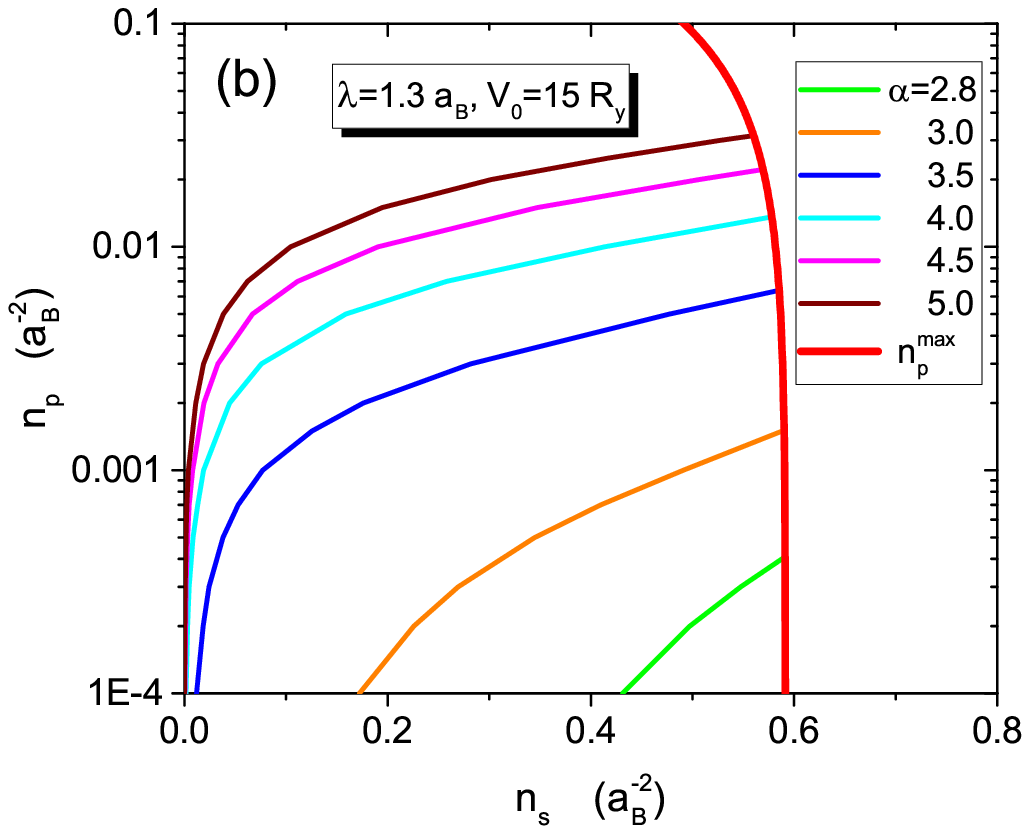}
      \includegraphics[scale=0.67]{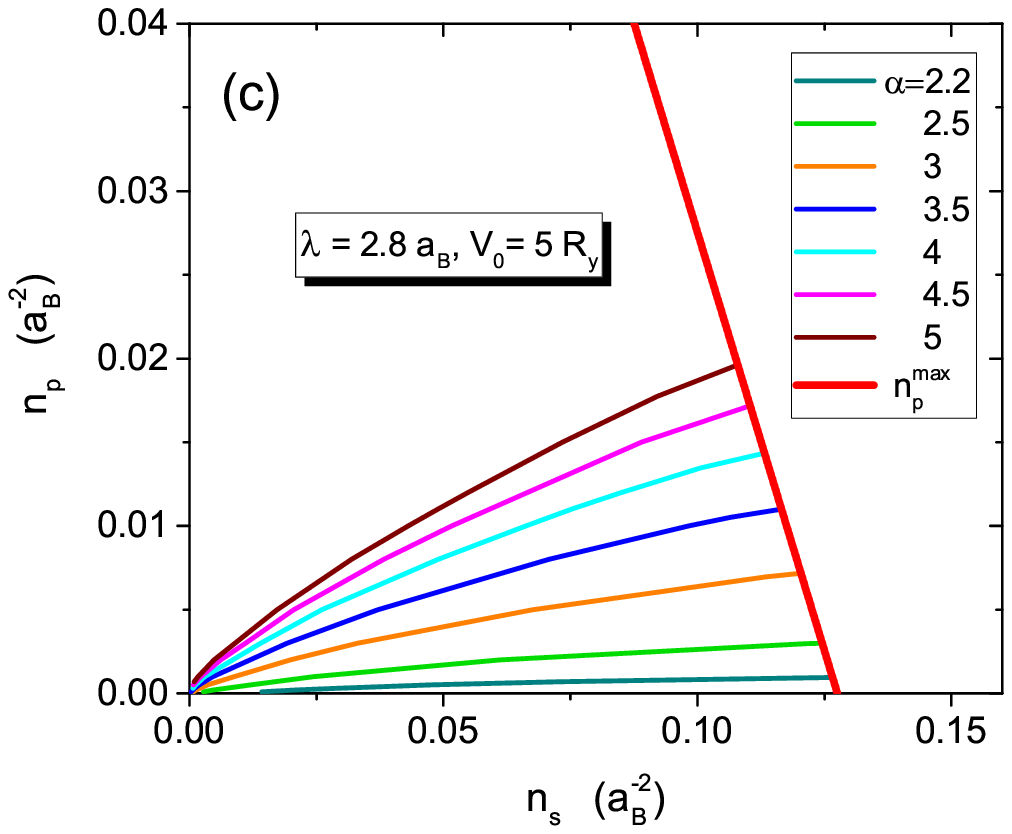}
       \caption{The stabilized electron-pair density $n_p$ in (a) linear scale and (b) logarithmic scale
       as a function of the single-electron density $n_s$ in the system with $\lambda=1.3$ a$_B$ and
       $V_0=$15 R$_y$ for the e-ph coupling constant values $\alpha=2.8$ to 5.0. (c) The same as (a) but
       for $\lambda=2.8$ a$_B$, $V_0=$5 R$_y$ and $\alpha=2.2$ to 5.0. The thick-red curve indicates
       the maximum allowed electron-pair density $n^{\rm max}_{p} =\lambda^{-2}-n_{s}$.}
       \label{PP1}
\end{figure}

Fig.~\ref{Ealfa}(a) shows for fixed total electron density $n_t = 0.5$ a$_B^{-2}$ and e-ph coupling
constant $\alpha=4$, the dependence of the single-electron band bottom $E^{\Gamma}_{\rm s}$,
its Fermi energy $E_F$, and the electron-pair band bottom $E^{\Gamma}_{\rm p}$ on the electron
distribution in the two bands $n_s$ and $n_p$. The total energy per electron $E_{\rm tot}$ is also given
in the figure. It is the average ground-state energy per electron in the system calculated by Eqs.~(\ref{EGS}),
(\ref{EGSel}), and (\ref{Epol}), and is given by
\begin{eqnarray}
 E_{\rm tot} &=& \frac{E_{\rm GS}}{N_t} = \frac{E^{\rm (el)}_{\rm GS} + E^{\rm (tot)}_{\rm pol} }{N_t}
  \nonumber \\
 &=& \frac{n_s}{n_t} ( E^{\Gamma}_{\rm s}+\frac{\varepsilon_F}{2})+ 2 \frac{n_p}{n_t} E^{\Gamma}_{\rm p},
\end{eqnarray}
where $E^{\Gamma}_{\rm s}$ and $E^{\Gamma}_{\rm p}$ are given by Eqs.~(55) and (56), respectively.
It is seen that, though the total electron density $n_t$ is a constant, the above obtained energies
depend on the distribution of the electrons between the two bands. Especially the energy of the electron-pair
band $E^{\Gamma}_{\rm p}$ depends strongly on $n_p$.
In Fig.~\ref{Ealfa}(a) we see that with increasing the single-electron density $n_s$,
the single-electron band $E^{\Gamma}_{\rm s}$ and the Fermi energy $E_F$ vary slowly.
Because $n_t$ is a constant, increasing $n_s$ means decreasing $n_p$.
At low density $n_p$, the weak screening in the electron-pair band enhances greatly the bipolaron energy
resulting in a lowering of the bottom of the electron-pair band $E^{\Gamma}_{\rm p}$ which
reaches the Fermi surface.
We find that under the considered condition, $E^{\Gamma}_{\rm p}$ touches the Fermi surface
at $n_s=0.4774$ $a_B^{-2}$ and $n_p=0.0113$ $a_B^{-2}$ as indicated by the black dot.
For these densities, the electron pairs (or the bipolarons) are stabilized on the Fermi surface.
It means that only 4.5\% of the electrons in the system form stable electron pairs in this case.
If we look at the total energy $E_{\rm tot}$, it tends to reduce the energy of the system at
the higher single-electron density side. But the electron pairs on the Fermi surface cannot
decay into two single electrons due to the Pauli exclusion principle. Moreover,
breaking an electron pair needs a cost to overcome their correlation energy.
Therefore, the electron-pair density is stabilized at $n_p=0.0113$ $a_B^{-2}$ in the ground state
of the system with $\alpha=4$ and $n_t=0.5$ $a_B^{-2}$.
Fig.~\ref{Ealfa}(b) shows the electron-pair band energy $E^{\Gamma}_{\rm p}$ and the
single-electron band Fermi energy $E_F$ in the same system with $n_t=0.5$ $a_B^{-2}$
but for different e-ph coupling constant $\alpha$.
We see that for $\alpha=2$ no stable electron pairs are found. The calculations indicate that
for $\alpha \gtrsim 3$, part of the electrons can form electron pairs on the Fermi surface
at low electron-pair density. The stabilized electron-pair densities are $n_p= 0.001$, 0.011, and
0.026 a$^{-2}_B$ for $\alpha=3$, 4 and 5, respectively.

In order to determine the electron-pair density in the ground-state of the system,
we solve the following equation
\begin{equation}
   E^{\Gamma}_{\rm p} (n_s,n_p) = E_F(n_s,n_p),
\end{equation}
as a function of $n_s$ and $n_p$ for fixed $\lambda$, $V_0$ and $\alpha$.
Figs.~\ref{PP1}(a) and \ref{PP1}(b) show the density of stabilized electron-pairs or bipolarons
in linear and logarithmic scale, respectively, as a
function of the single-electron density in the system with $\lambda=1.3$ a$_B$ and $V_0=$15 R$_y$
for $\alpha=2.8$, 3.0, 3.5, 4.0, 4.5, and 5.0. We see that for $\alpha=2.8$ a very low density
of electron pairs become stabilized. With increasing $\alpha$, more electron pairs appear in the system.
For $\alpha=5$, their density may reach $n_p$=0.032 a$^{-2}_B$. This corresponds to about
11\% of the electrons in the system are in the electron-pair band.
The thick-red curve in the figure is the possible maximum
electron-pair density $n^{\rm max}_{p}$ given by Eq.~(\ref{npmax}).
It is important to notice that the relation between $n^{\rm max}_{p}$ and $n_s$ not only
restrict the electron-pair density being less than $n^{\rm max}_{p}$. It also means
that if $n_s$ is larger than $\lambda^{-2}$, i.e., the single-electron band filling is more
than half, there are no stable electron pairs in the system.
For the crystal with $\lambda=1.3$ a$_B$, the half filling of the single-electron band occurs at
$n_s=0.5917$ a$^{-2}_B$.

In Fig.~\ref{PP1}(c) we show the electron-pair density in a different 2D potential with
$\lambda=2.8$ a$_B$ and $V_0=$5 R$_y$. In this case, the band-gap $E_g^{(0)}$
and also $\nu_0$ are small, as shown in Fig.~2. The half filling of the single-electron
band is at $n_s=0.1276$ a$^{-2}_B$. We find that for $\alpha \simeq 2$, some electron
pairs can already be stabilized with small density. For $\alpha=5$, the electron-pair
density may reach 0.02 a$^{-2}_B$ corresponding to 31\% of the total electrons
being in the paired state.

\section{Summary and outlook}

The starting point of our work is that the electron-electron correlation
is orbital dependent. The exchange-correlation energy of an electron pair occupying the same
orbital is larger than the average of the exchange-correlation energy of many electrons.
Depending upon the crystal structure and potential, such electron pairs can form a metastable
electron-pair band. The metastable electron-pair band is obtained in two different ways.
One follows the tight-binding method and the other is similar to the nearly-free electron model.
They give consistent results for the dispersion relation of the electron-pair states
in the period potential.

Combining the metastable electron-pair band together with the corresponding single-electron band,
we constructed a many-particle system consisting of single electrons and metastable electron pairs.
When further considering many-body interaction renormalization including single electrons, electron
pairs, and optical phonons, we found that the metastable electron pairs can be stabilized.
The calculations show that the polaron effects play an essential role in counterbalancing
the Coulomb repulsion in the stabilization of the electron pairs. The electron-phonon coupling
is enhanced in the strongly correlated electron pairs and leads to the formation of bipolarons.
On the other hand, screening affects significantly both the polaron and bipolaron energies
manifesting a cooperative interplay of electron-electron and electron-phonon interactions.
The obtained ground state of the system consists of polarons in the single-electron band
and bipolarons in the electron-pair band sitting on the Fermi surface of the single-electron band.

The numerical calculations presented in this paper are performed for simple potentials and with
the $s$-orbital in 2D square-lattice crystals. However, the physical processes and numerical
calculations for electron pairing can be extended to a quasi-2D crystal of a single atomic layer
with finite thickness or to 3D systems.
In principle, it can also be extended to study the electron pairing in $p$- and $d$-orbitals.
The obtained results within the present simple model predict light and small electron
pairs (bipolarons) in the crystal. In the center-of-mass coordinates, the electron pair is
a Bloch wavefunction with an effective mass of twice a single electron.
The pair is small and local because the average separation between the two electrons is less than
half of the lattice constant and they are localized in the same unit cell when expressed in the
relative coordinates. Furthermore, only a fraction of the electrons in the system form pairs.

We have obtained a many-particle ground state with spin-singlet electron pairs in the condensate phase.
We expect that these preformed electron pairs at certain densities in coherent state will contribute
to superconductivity.
Finally, we want to comment on the possible ``superconducting energy gap". From our calculations
we naturally infer such a gap to the transition energy required to break the stabilized electron pairs 
sitting on the Fermi surface. 
This transition is determined by the Hamiltonian ${H}^{\rm t}_{\rm int}$ in Eq.~(\ref{Hintt})
where the potential $v^{\rm t}(q)$ was given by Eq.~(11) in Ref.~[\onlinecite{hai2015}].
Because the electron pairs are stabilized on the Fermi surface in the center of the Brillouin
zone at ${\bf k}=0$, an external excitation has to overcome the internal correlation energy of an electron pair
to bring two electrons to the single-electron states above the Fermi surface at $\pm {\bf k} \neq 0$.
The minimum energy required (or the gap) is primarily determined by the potential $v^{\rm t}(q)$
and the final momenta $\pm \hbar {\bf k}$ of the single-electron states. It is band structure dependent, anisotropic, and nodeless.

\acknowledgments
{This work was supported by the Brazilian agencies FAPESP and CNPq. GQH would like to thank Prof. Bangfen Zhu for
his invaluable support and expert advice.}

\end{document}